\documentclass[usenatbib,fleqn]{mnras}
\usepackage{graphicx}
\graphicspath{{./plots/}}
\usepackage{float}
\usepackage{amssymb,amsmath}
\usepackage{color}
\usepackage{multirow}
\usepackage{soul}
\usepackage{orcidlink}

\newcommand{\sarc}{$^{\prime\prime}\!\!$}

\newcommand{\tbpeak}{$T_{b,\mathrm{peak}}$}
\newcommand{\tbtotal}{$T_{b,\mathrm{total}}$}

\renewcommand{\aa}{A\&A}

\renewcommand{\aj}{AJ}

\renewcommand{\apj}{ApJ}
\renewcommand{\apjs}{ApJS}
\renewcommand{\araa}{ARA\&A}

\renewcommand{\mnras}{MNRAS}
\renewcommand{\nat}{Nature}

\renewcommand{\pasp}{PASP}

\renewcommand{\apjl}{ApJL}
\newcommand{\ascl}{ASCL}
\newcommand{\natast}{Nature Astronomy}

\title[AGN identification with LOFAR]{Identifying active galactic nuclei via brightness temperature with sub-arcsecond International LOFAR Telescope observations}
\author[Leah K. Morabito]{\parbox{\textwidth}{Leah K. Morabito$^{1,2}$\thanks{E-mail: leah.k.morabito@durham.ac.uk}\orcidlink{0000-0003-0487-6651},
F. Sweijen$^{3}$,
J. F. Radcliffe$^{4,5}$, 
P.N. Best,$^{6}$,
Rohit Kondapally$^{6}$
Marco Bondi$^{7}$,
Matteo Bonato$^{7,8}$
K. J. Duncan$^{6}$,
Isabella Prandoni$^{7}$,
T. W. Shimwell$^{3,9}$,
W. L. Williams$^{3}$,
R.J. van Weeren$^{3}$,
J. E. Conway$^{10}$, 
G. Calistro Rivera$^{11}$ \\}\\ 
$^{1}$Centre for Extragalactic Astronomy, Department of Physics, Durham University, Durham DH1 3LE, UK \\
$^{2}$Institute for Computational Cosmology, Department of Physics, University of Durham, South Road, Durham DH1 3LE, UK \\
$^{3}$Leiden Observatory, Leiden University, P.O. Box 9513, 2300 RA Leiden, The Netherlands \\
$^{4}$Department of Physics, University of Pretoria, Lynnwood Road, Hatfield, Pretoria 0083, South Africa \\
$^{5}$Jodrell Bank Centre for Astrophysics, University of Manchester, Oxford Road, Manchester M13 9PL, UK \\
$^{6}$Institute for Astronomy, Royal Observatory, Blackford Hill, Edinburgh, EH9 3HJ, UK \\
$^{7}$INAF - Istituto di Radioastronomia, Via P. Gobetti 101, 40129, Bologna, Italy \\
$^{8}$Italian ALMA Regional Centre, Via Gobetti 101, I-40129, Bologna, Italy \\
$^{9}$ASTRON, Netherlands Institute for Radio Astronomy, Oude Hoogeveensedijk 4, 7991 PD, Dwingeloo, the Netherlands \\
$^{10}$Department of Space, Earth and Environment, Chalmers University of Technology, Onsala Space Observatory, SE-439 92 Onsala, Sweden \\
$^{11}$European Southern Observatory, Karl-Schwarzschild-Stra\ss e 2, 85748 Garching bei M\"{u}nchen, Germany \\ }

\begin{document}
\date{}
\pagerange{\pageref{firstpage}--\pageref{lastpage}} \pubyear{2021}
\maketitle

\label{firstpage}

\begin{abstract}
Identifying active galactic nuclei (AGN) and isolating their contribution to a galaxy's energy budget is crucial for studying the co-evolution of AGN and their host galaxies. Brightness temperature ($T_b$) measurements from high-resolution radio observations at GHz frequencies are widely used to identify AGN. Here we investigate using new sub-arcsecond imaging at 144 MHz with the International LOFAR Telescope to identify AGN using $T_b$ in the Lockman Hole field. We use ancillary data to validate the 940 AGN identifications, finding 83 percent of sources have AGN classifications from SED fitting and/or photometric identifications, yielding 160 new AGN identifications. Considering the multi-wavelength classifications, brightness temperature criteria select over half of radio-excess sources, 32 percent of sources classified as radio-quiet AGN, and 20 percent of sources classified as star-forming galaxies. Infrared colour-colour plots and comparison with what we would expect to detect based on peak brightness in 6\sarc\ LOFAR maps, imply that the star-forming galaxies and sources at low flux densities have a mixture of star-formation and AGN activity. We separate the radio emission from star-formation and AGN in unresolved, $T_b$-identified AGN with no significant radio excess and find the AGN comprises $0.49\pm 0.16$ of the radio luminosity. Overall the non-radio excess AGN show evidence for having a variety of different radio emission mechanisms, which can provide different pathways for AGN and galaxy co-evolution. This validation of AGN identification using brightness temperature at low frequencies opens the possibility for securely selecting AGN samples where ancillary data is inadequate. 
\end{abstract}

\begin{keywords}
galaxies: active -- galaxies: jets -- radio continuum: galaxies -- acceleration of particles -- radiation mechanisms: non-thermal
\end{keywords}

\section{Introduction}
\label{sec:intro}

Understanding how active galactic nuclei (AGN) interact with their host galaxies is one of the main challenges in astronomy today. The fact that some sort of feedback links the growth of galaxies to the super-massive black hole which powers the AGN is widely accepted. Tight empirical relations between observed host galaxy properties \citep[e.g., velocity dispersion, bulge mass;][]{ferrarese_fundamental_2000,gebhardt_relationship_2000} and the mass of the central super-massive black hole are a strong, but indirect, argument for feedback. Cosmological simulations require AGN feedback to suppress the growth of the most massive galaxies and reproduce the stellar mass function we observe in the Universe today \citep[e.g.,][]{bower_breaking_2006,croton_many_2006}. Although it is clear from both observations and theory that AGN feedback is an important component in a galaxy's growth, we do not yet understand the details of how AGN feedback works. 

A significant limitation in understanding how AGN feedback works is what it operates on: star formation. AGN feedback can stimulate star formation \citep[e.g.,][]{de_young_emission_1981,silk_unleashing_2013,zhuang_black_2021} by creating the dense, turbulent gas conditions which lead to star formation. Feedback could also  suppress / quench star formation by heating or removing the gas from which stars would form \citep[see, e.g.][]{bower_breaking_2006,greene_feedback_2011,king_powerful_2015}. Either way, it can be difficult to pick apart observational evidence to understand which is happening (Ward et al., MNRAS accepted). Historically, radio-loud sources have provided excellent evidence for AGN feedback, particularly on large scales. High redshift radio-loud AGN exhibit an alignment between the optical and radio axes, implying that the radio jets may be stimulating star formation \citep{chambers_alignment_1987,best_evolution_1996,dey_triggered_1997,nesvadba_gas_2020}. In more local radio-loud AGN in the centres of clusters, we observe a balance between the radio jet power and the work it would take to create the co-spatial cavities in the hot gas seen in X-ray observations \cite[e.g.,][]{mcnamara_chandra_2000,birzan_systematic_2004,wise_x-ray_2007}. Galaxy-scale radio jets can also have internal lobe energy comparable to the energy in the interstellar medium  \citep[ISM;][]{webster_population_2021}, indicating that the ability of radio jets to impact a galaxy's evolution extends down to smaller scales. 

While it is easy to identify large-scale radio jets, once on sub-galactic scales the problem compounds: both AGN activity and star formation generate radio emission, making it difficult to understand their interplay. It is also not clear what the source of radio emission is in radio-quiet AGN; it may be due to small-scale jets, winds, or even star formation \citep[for a review, see][]{panessa_origin_2019}. The jury is even still out on whether the radio emission in radio-quiet AGN is linked to AGN activity \citep[e.g.,][]{white_radio-quiet_2015,zakamska_quasar_2014,macfarlane_radio_2021} or star formation \citep[e.g.,][]{condon_active_2013,padovani_radio-faint_2015,gurkan_lofar/h-atlas:_2018,radcliffe_radio_2021}. 
Multi-wavelength observations can help identify sources with radio excess above that expected from star formation \citep[e.g.,][]{wilson_star_1988,drake_radio-excess_2003,del_moro_goods-herschel_2013,calistro_rivera_lofar_2017}, but many AGN lie on the radio to far-infrared correlation \citep[e.g.,][]{sopp_composite_1991,bonzini_star_2015} and cannot be identified in this way.

The only way to unambiguously identify radio emission from AGN activity and simultaneously separate it from star formation is via brightness temperature, $T_b$. This is defined as the temperature of the blackbody which would produce the observed surface brightness (flux density per solid angle) at the observed radio frequency. There is a limit to the amount of radio flux density per solid angle that can be generated by star formation, even in the most luminous starburst galaxies \citep{condon_radio_1992} and particularly at redshifts $z>0.1$ \citep{kewley_compact_2000}. Therefore a value of brightness temperature above this limit provides a secure AGN classification, although the converse is not true: we cannot securely say there is no AGN in sources below this brightness temperature limit, although it is likely they are dominated by star formation. 

Brightness temperature measurements are often used in very long baseline interferometry (VLBI) experiments, where the high resolution can measure the appropriate parameter space to distinguish between star formation and AGN activity. At GHz frequencies, wide-field VLBI has shifted from processing the entire primary beam for a single correlated phase centre \citep{garrett_agn_2001} to the more computationally feasible processing of multiple phase centres correlated independently across a larger field of view \citep{deller_difx-2:_2011,morgan_vlbi_2011}. This has allowed surveys to follow up radio sources with brightness temperature measurements in targeted observations of select fields \citep[e.g.,][]{muxlow_high-resolution_2005,chi_deep_2013,radcliffe_nowhere_2018}, up to areas of $\sim$2 deg$^2$ \citep{herrera_ruiz_faint_2017}. However, the limiting value of $T_b$ depends not only on resolution but also frequency (see \S~\ref{sec:tb}).

The International LOFAR Telescope \citep[ILT;][]{van_haarlem_lofar:_2013} is a phased array operating at 10 -- 240 MHz, with stations spread across Europe. With baselines up to $\sim$2,000 km, the ILT can achieve sub-arcsecond resolution at MHz frequencies. Sub-arcsecond imaging with the full ILT shares many common challenges with high-frequency VLBI, but with the addition of its own unique challenges at low frequencies \citep[for more details, see][]{morabito_sub-arcsecond_2022}. Imaging of individual sources has yielded significant results \citep[e.g.,][]{varenius_subarcsecond_2015,morabito_lofar_2016,ramirez-olivencia_sub-arcsecond_2018,timmerman_origin_2022,kukreti_unmasking_2022,groeneveld_pushing_2022}, but these studies have remained in the realm of bright sources which can be self-calibrated. The real power of the ILT is in its wide field of view, allowing access to fainter sources. \cite{sweijen_deep_2022} recently published the first full field of view image covering 6.6 deg$^2$ at the full resolution of the ILT. In this $\sim$7 billion pixel image, there are 2,316 sources detected with 5$\sigma$ significance (after removing duplicate sources). 

In this paper we exploit this ILT dataset to study the brightness temperatures of the low frequency radio population, for the first time. We begin by describing the data in Section~\ref{sec:data}. Section~\ref{sec:tb} opens with a discussion on brightness temperature at low frequencies, followed by the sample selection. Results are presented in Section~\ref{sec:results} followed by the separation of star formation and AGN in Section~\ref{sec:sfagn}. Discussion and conclusions are in Sections~\ref{sec:discussion} and \ref{sec:con}, respectively. Throughout the paper, we assume the \textsc{WMAP9} cosmology \citep{hinshaw_nine-year_2013} in {\tt astropy} and define radio flux density as $S_{\nu}\propto \nu^{\alpha}$, where $\alpha$ is the radio spectral index. The code which generated the plots and results is publicly available at \url{https://github.com/lmorabit/ILT_AGNdetect}.

\section{Data}
\label{sec:data}

\subsection{Lockman Hole data}
We use the catalogue from \cite{sweijen_deep_2022}. This catalogue is drawn from the first high-resolution (0\sarc .3$\times$0\sarc .4) wide-field image made with LOFAR. This image, which covers 6.6 deg$^2$ of the Lockman Hole, has a median rms noise level (across the rms map) of 34$\,\mu$Jy$\,$beam$^{-1}$. We will refer to this image as the \textit{high-resolution image}. We refer the reader to \cite{sweijen_deep_2022} for full details on how this image was generated, but briefly summarise relevant information here. 

The wide-field image is made up of 25 `facets' following the technique developed by \cite{van_weeren_lofar_2016,van_weeren_lofar_2021}. Corrections for the point spread function (PSF) and astrometry were applied per facet, and a final flux density scale correction applied for the whole field. The Python Blob Detector and Source Finder \citep[\textsc{PyBDSF};][]{mohan_pybdsf:_2015} was run individually on the 25 facet images before the outputs were combined into a single catalogue, where duplicate sources were removed. Removal of duplicate sources was done first using simple positional matching and followed up with visual identification. The resulting catalogue contains 2,316 unique sources where the peak brightness, $S_p$, is  $\geq 5\sigma$ (using the rms noise for the associated island). The median rms noise in the full image is 34$\,\mu$Jy$\,$beam$^{-1}$, although this varies radially and is lower in the centre. 

This high-resolution catalogue was matched to the 6\sarc\ resolution image made as part of the LOFAR Surveys Deep Fields Data Release 1 \citep{tasse_lofar_2021}. We will refer to this image as the \textit{standard-resolution image}.  All of the sources detected at 5$\sigma$ in the high-resolution image have a match in the standard-resolution image, which is unsurprising as it uses $\sim$100 hours of data to reach $\lesssim 23\,\mu$Jy$\,$beam$^{-1}$ in the inner part of the field, rather than the 8 hours used for the high-resolution image. This provides information on angular scales $\geq$20 times larger than the high-resolution catalogue, and in particular, is more sensitive to diffuse emission. 

As part of the Deep Fields Data Release 1, science-ready catalogues were produced that provide extensive information at infrared (IR) through X-ray bands. \cite{kondapally_lofar_2021} present the identification of counterparts with radio sources, and the careful compilation of available multi-wavelength information, for 31,162 sources in the standard-resolution image of the Lockman Hole field. This includes some flags for the identification of AGN from optical, IR, and X-ray information. The catalogues also include photometric redshifts from \cite{duncan_lofar_2021}, as only $\sim10$ percent of sources have spectroscopic redshifts. For the analysis presented in this paper, we remove sources which have no redshift information available ($<$5 percent), leaving 2,214 sources from the high-resolution image. 

\subsection{Parameters from SED fitting}
Source classifications, stellar masses, and star formation rates were calculated by Best et al., in prep. using SED fitting from the far-IR to the ultraviolet (UV). The SED fitting was carried out on all galaxies detected in the standard-resolution image using four different SED fitting software. The first two, \textsc{MAGPHYS} \citep{da_cunha_simple_2008} and \textsc{BAGPIPES} \citep{carnall_inferring_2018,carnall_how_2019}, operate on an energy balance approach, where the energy absorbed by dust in the optical and UV bands must match the thermal emission from dust in the far-IR through sub-mm. While these two codes generally give consistent results for high signal-to-noise data, they do not include any AGN component. \textsc{CIGALE} \citep{burgarella_star_2005,noll_analysis_2009,boquien_cigale_2019} also operates on the energy balance principle, and includes an AGN component. The final code, \textsc{AGNfitter} \citep{calistro_rivera_agnfitter_2016}, is designed to work in cases where the assumption of energy balance may not hold if the UV light and IR emission are spatially distinct. This is often the case for AGN. 

Best et al., in prep. modelled all of the galaxies using each of the four codes, including twice for \textsc{CIGALE}, using a different suite of models for the AGN component each time. This provided radiative mode AGN identifications. Radio AGN were identified using the relationship between radio luminosity at 150 MHz, $L_{150}$, and the derived SFRs. Best et al., in prep. defined radio-excess identified AGN as lying further than 0.7 dex from the ridge line of this relationship.  For all fitted galaxies, the consensus stellar masses and SFRs were derived (Section 5; Best et al. in prep) using a Chabrier initial mass function \citep{chabrier_galactic_2003}.

\section{Sample selection}
\label{sec:tb}

\subsection{Brightness temperatures at low frequencies}

Frequently, in the literature one will see that a source is reliably considered to be an AGN if it exceeds $T_b > 10^5\,$K. However, the exact limiting value of $T_b$ depends on the observing frequency, non-thermal spectral index, the redshift, and the temperature of the gas. It is worth revisiting this in light of these new LOFAR observations.

We start from brightness temperature, which is defined (in the Rayleigh-Jeans approximation) as: 
\begin{equation}
\label{eqn:tb}
T_b =  \frac{c^2}{2 k_{b}} \frac{S_\nu}{\nu^2\Omega} \,\textrm{K},
\end{equation}
where $c$ is the speed of light, $k_B$ is the Boltzmann constant, $S_{\nu}$ is the flux density, $\nu$ is the observing frequency, and $\Omega$ is the source solid angle. For a Gaussian source, 
\begin{equation}
\label{eqn:solidangle}
\Omega = \frac{\pi \theta_1 \theta_2}{4\textrm{ln}2},
\end{equation}
where $\theta_1$, $\theta_2$ are the fitted Major and Minor axes of a 2D Gaussian. Rearranging Eqn.~\ref{eqn:tb}, we arrive at:
\begin{equation}
\label{eqn:fluxperSA}
\frac{S_{\nu}}{\Omega} = \frac{T_b \nu^2}{1.38\times10^{24}},
\end{equation}
in units of Jy$\,$arcsec$^{-2}$ (flux density per solid angle)\footnote{Note that the solid angle as defined in Eqn.~\ref{eqn:solidangle} is rarely explicitly used; a commonly seen version of Eqn~\ref{eqn:fluxperSA} will absorb the constants and use $\theta_1 \theta_2$ directly, resulting in a constant of $1.22\times10^{24}$ instead.}.

Starting from the simplest case, the brightness temperature from a normal star-forming galaxy \citep{condon_radio_1992} is given by: 
\begin{align}
\label{eqn:galtb}
&T_b = T_e [1 - \textrm{exp}(-\tau)]\left( 1 + 10\left(\frac{\nu}{\textrm{GHz}}\right)^{0.1+\alpha}\right), \\
&\textrm{with}\, \tau = \tau_0 \left(\frac{\nu}{\nu_0}\right)^{-2.1} \notag,
\end{align}
where $\tau_0=1$ at frequency $\nu_0$. 
In Eqn.~\ref{eqn:galtb} we can see that the brightness temperature depends on the frequency at which the optical depth reaches unity, $\nu_0$, and the electron temperature of the gas, $T_e$. It also depends on the synchrotron spectral index, $\alpha$. For this initial study we choose to use this simple model which assumes a single-phase ISM, and build on this in future work. 

\begin{figure*}
\centering
\includegraphics[width=0.45\textwidth]{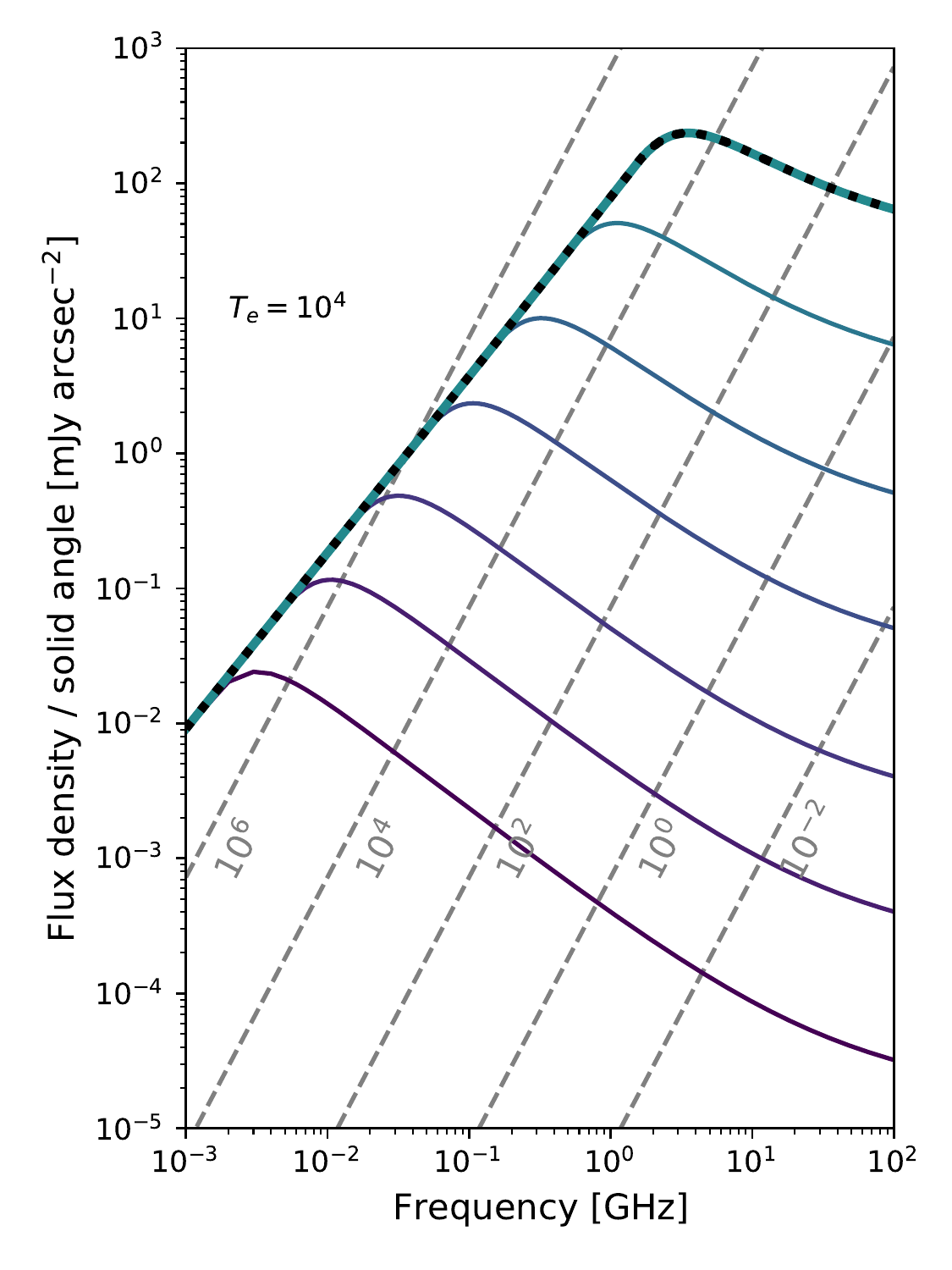}
\includegraphics[width=0.45\textwidth]{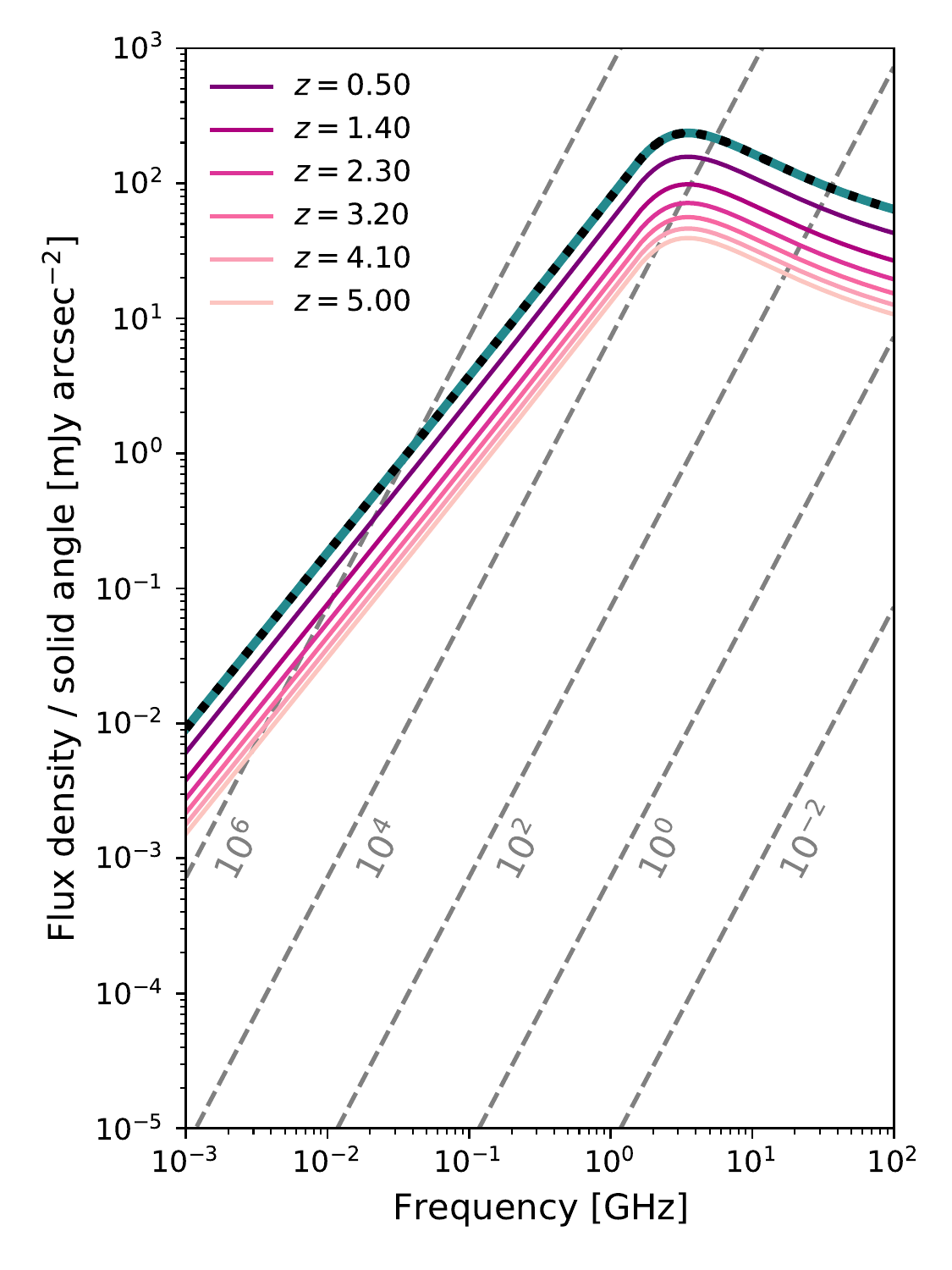}
\includegraphics[width=0.45\textwidth]{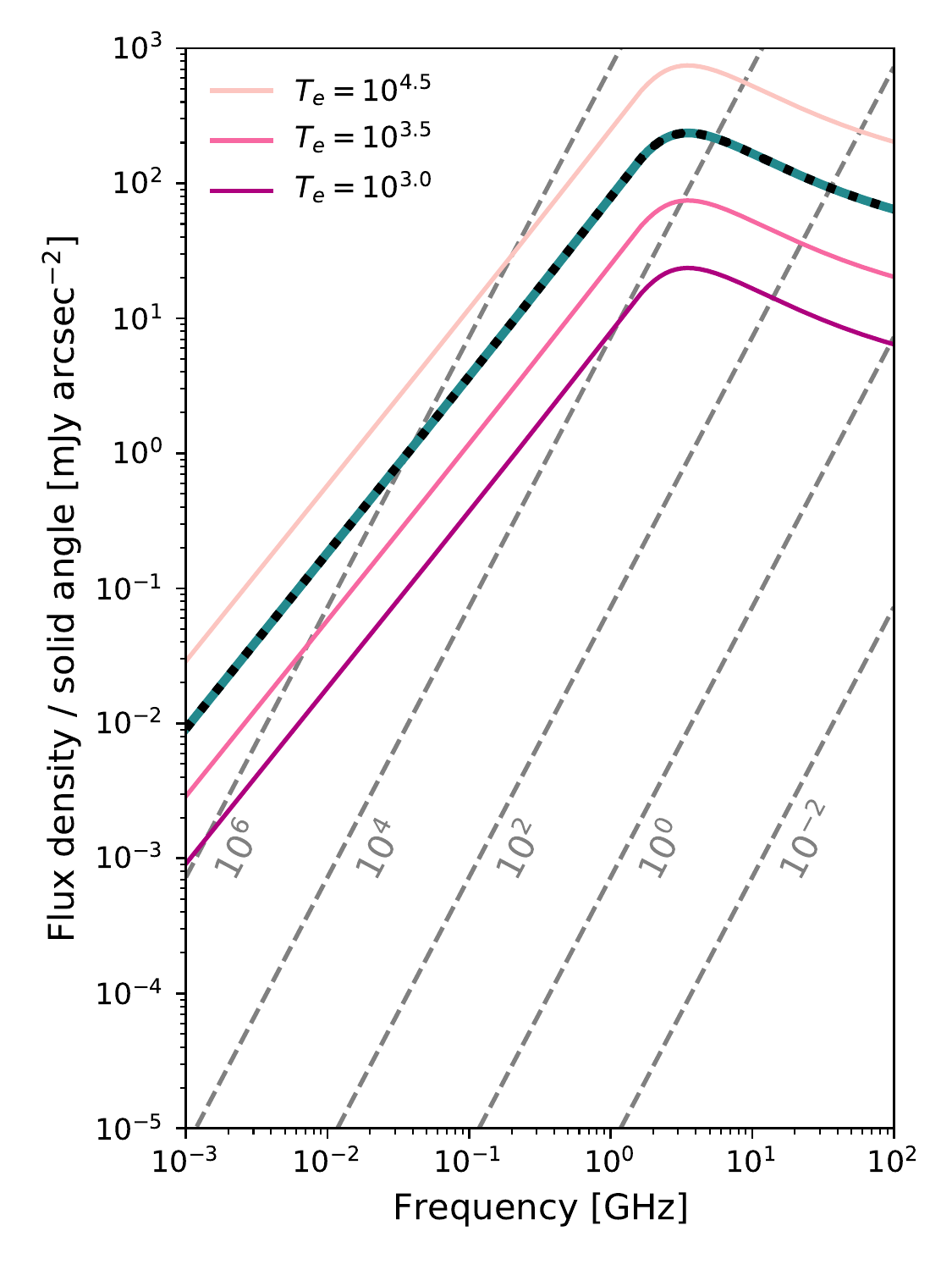}
\includegraphics[width=0.45\textwidth]{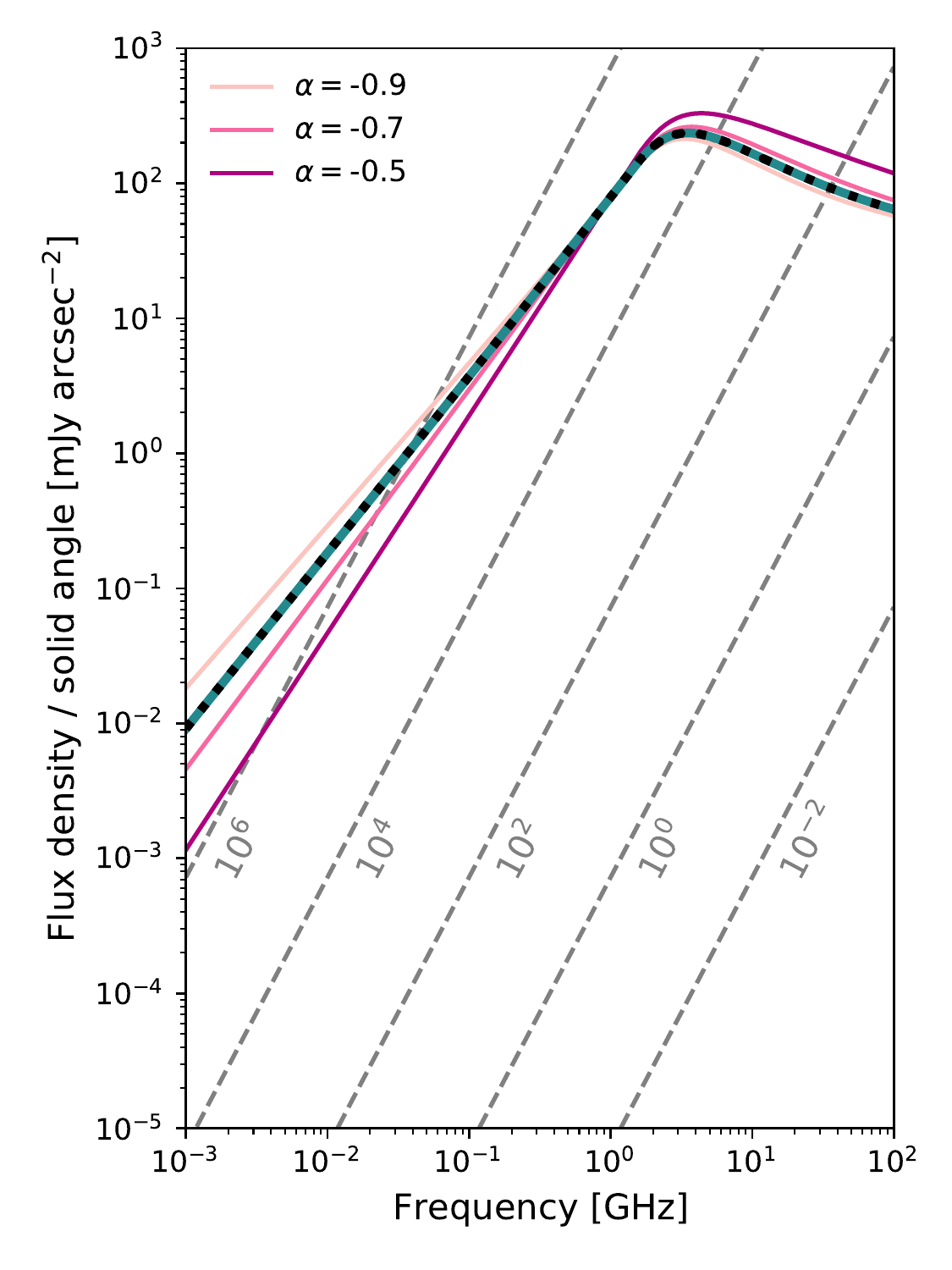}
\caption{\label{fig:condon} The top left panel shows flux density per solid angle versus frequency \citep[reproduction of Fig. 4 in][]{condon_radio_1992}. The dashed gray lines show constant brightness temperatures, as labelled in the plot. The coloured lines show the limiting flux per solid angle as a function of frequency, for $\nu_0$ (the frequency at which the optical depth is unity) of 3 MHz, 10 MHz, 30 MHz, 100 MHz, 300 MHz, 1 GHz, and 3 GHz (from purple to green, bottom to top). The line with $\nu_0=$3 GHz is thicker in width and marked with black dots, and this is plotted in the other three panels for reference.  For $\nu_0 = 3\,$GHz, the top right panel shows how this changes with redshift. The bottom left panel shows how it changes with electron temperature, $T_e$, while the bottom right panel shows the change with spectral index, $\alpha$. In all figures the $z=0$, $T_e=10^4\,$K, $\alpha=-0.8$ line is shown as the thicker line dotted with black.}
\end{figure*}

Inserting Eqn.~\ref{eqn:galtb} into Eqn.~\ref{eqn:fluxperSA} yields a beam-independent model of flux per solid angle, and we arrive at Figure~\ref{fig:condon}, which shows the flux per solid angle versus observing frequency. The top left panel is a reproduction of Fig. 4 from \cite{condon_radio_1992}, with a range of values for $\nu_0$, for a typical electron temperature $T_e=10^4\,$K. As $\nu_0$ increases, the frequency at which the flux per solid angle peaks also increases. At an observed frequency of 1.4 GHz, the maximum value allowed out of all of the curves is $\sim 10^5\,$K. Of course if $\nu_0$ causes the flux per solid angle to peak at a lower frequency, the limiting value of $T_b$ will drop.  As the value of $\nu_0$ is virtually impossible to measure accurately for a distant galaxy, the most restrictive curve (i.e., $\nu_0=$3$\,$GHz) must be assumed for secure AGN identification. Assuming this most restrictive value of $\nu_0$, the three remaining panels of Fig.~\ref{fig:condon} show how this changes with redshift, electron temperature, and synchrotron spectral index. As the redshift increases, the curve shifts down, and therefore the limiting value of $T_b$ drops. Lower values of $T_e$ will also shift the curve down, while higher values shift it up. Synchrotron spectral index changes the shape of the curve, with steeper spectral indices reaching higher limiting values of $T_b$ at low frequencies (but lower at high frequencies).

It is important to remember that $T_b$ is not a physical temperature, and that the limiting value which divides star formation and AGN activity can change based on several parameters, e.g. with observed frequency: for $z=0$, $T_e=10^4\,$K the limiting value is $T_b \sim 10^5\,$K at 1.4 GHz, but closer to $T_b \sim 10^6\,$K at 144 MHz. The limiting value of $T_b$ should therefore be calculated per galaxy using all information available. For this work, we assume $T_e = 10^4\,$K, use the redshift of each individual source, and assume a spectral index of $\alpha = -0.8$.  This last assumption is conservative: we know that compact sources tend to have flatter spectral indices \citep[e.g.,][]{jackson_sub-arcsecond_2022}, which would lead to a lower limiting value of $T_b$ (see bottom right of Fig.~\ref{fig:condon}). In the absence of complementary spatially resolved spectral index information (the integrated spectral index in the standard resolution image may not be the same as the compact component), we opt to fix the spectral index for all sources. An ongoing survey with e-MERLIN and the EVN (PI: McKean) will, in the future, provide appropriate spectral index information to be able to measure the spectral index per source. This is more appropriate than comparison with a previous VLBA study \citep{middelberg_mosaiced_2013}, which is discussed in Appendix~\ref{app:vlba}.

\subsection{Selecting AGN using brightness temperature}
\label{subsec:tb}

We select AGN by calculating the limiting brightness temperature assuming $\nu_0=3\,$GHz, $T_e=10^4\,$K, $\alpha=-0.8$, and the specific redshift per source. We use spectroscopic redshifts where they exist, supplemented by photometric redshifts for the rest of the sample. Practically, we need the flux density information, $S$, and the area over which it is spread, $\Omega$.  For the area over which the flux density is spread, ideally we would use the real angular size of the source. This information is not directly available, so we use the deconvolved sizes from \textsc{PyBDSF}. Where the deconvolved major or minor axis is smaller than the limiting resolution (0\sarc .4 and 0\sarc .3, respectively), we replace the size with the limiting resolution \citep[see][]{lobanov_resolution_2005,radcliffe_nowhere_2018}. This will lead to an underestimation of $T_b$, and therefore we expect the AGN identifications to be secure. We treat the major and minor axes separately, and this impacts a total of 950 unique sources.

Using deconvolved sizes is not ideal, as the actual point spread function (PSF; as opposed to the restoring beam, which is the 2D Gaussian fit to the PSF used by \textsc{PyBDSF}) is the convolution of several kernels: \textit{$(i)$} the $u$-$v$ sampling function (dirty beam), \textit{$(ii)$} the auto-correlation function (ACF) of any residual calibration errors, and \textit{$(iii)$} the PSF distortions due to smearing. However, the fitted Gaussian restoring beam will be smaller than the actual PSF, meaning that when PyBSDF uses the restoring beam to calculate the deconvolved sizes they will therefore be an \textit{overestimate} of the true size, which leads to an \textit{underestimate} of the brightness temperature. The results in this paper are therefore conservative. To properly calculate the brightness temperature accounting for smearing and other intensity losses, one would need to simulate a dirty beam from the $u$-$v$ coverage of a particular observation, convert it to a residual image (this can be done as the dirty beam is the ACF of the noise, see \cite{schreiber_low_2021} Appendix A for details), compare that to the actual residuals to find the calibration error PSF, and finally calculate the smearing PSF. This procedure is computationally expensive, and we leave it to future work; the imaging strategy is still undergoing optimisation and the final procedure is not set in stone. For now, we note that our $T_b$ estimates are lower limits, and in the future we may be able to access more sources with high values of $T_b$. To be clear, this means \textit{we expect our current AGN identifications to be secure}, and we may be able to add more AGN identifications in the future. 

\begin{figure*}
    \centering
    \includegraphics[width=0.4\textwidth]{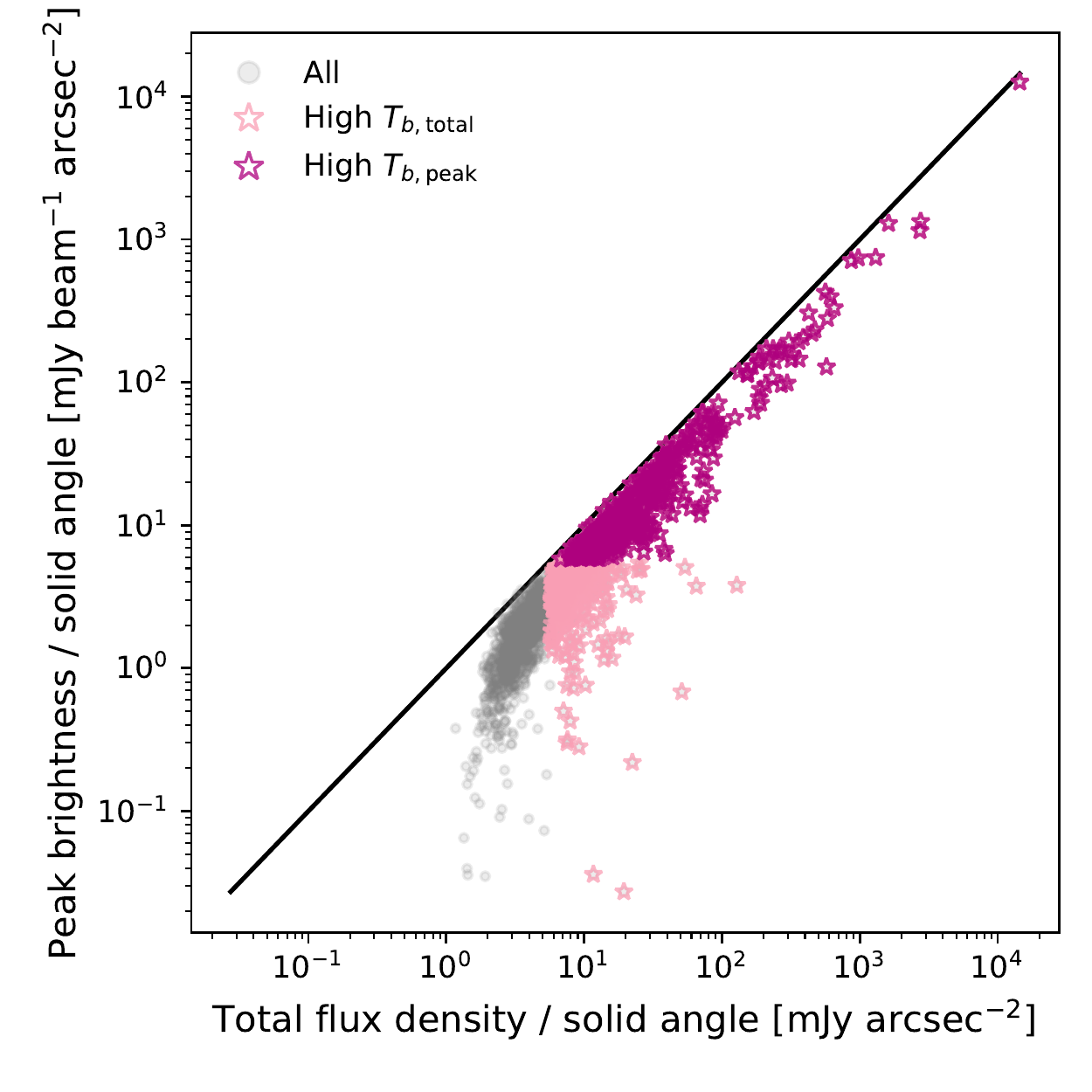}
    \includegraphics[width=0.45\textwidth]{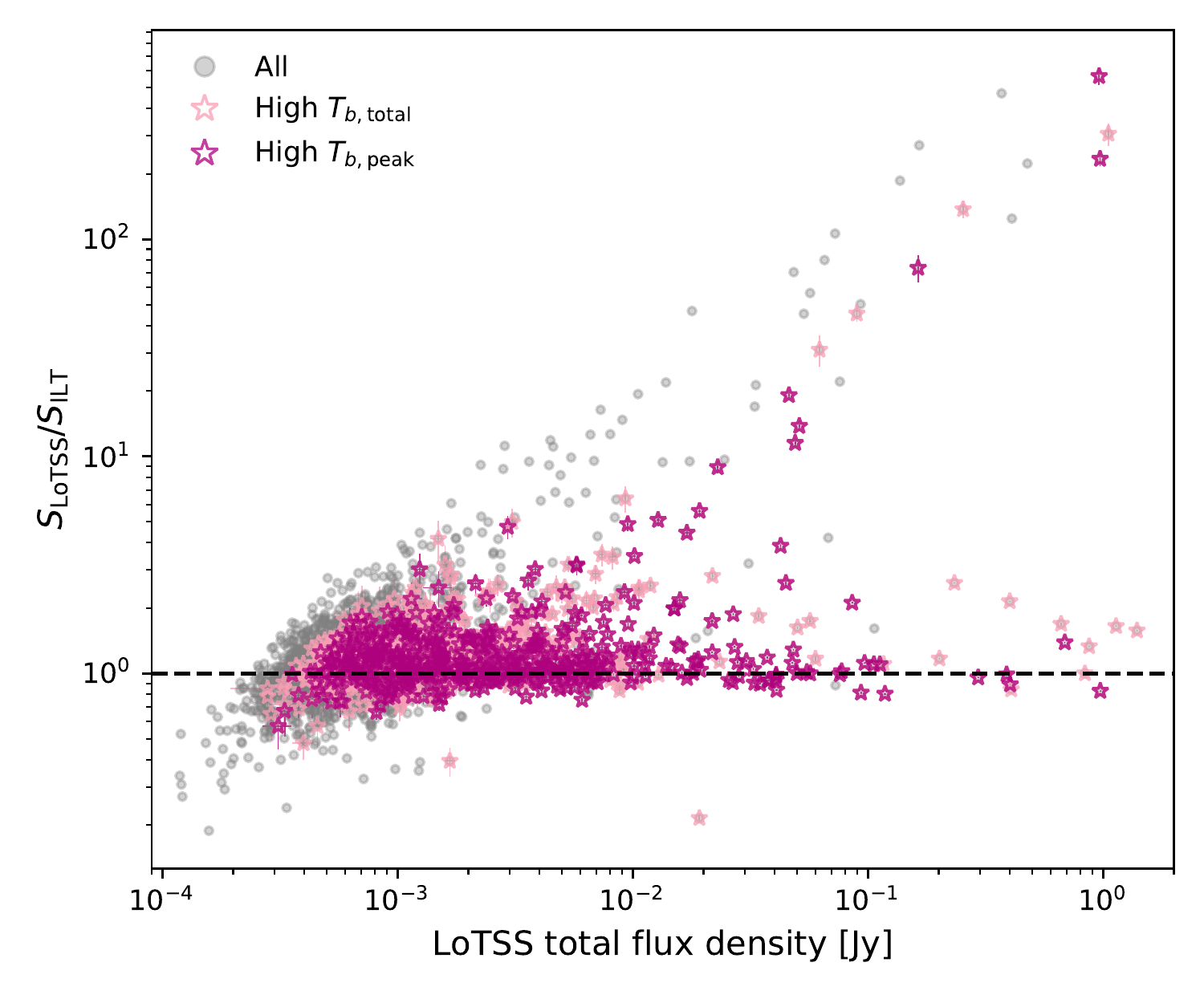}
    \caption{Observed properties of the high-$T_b$ selected sample. On the left, we show the peak brightness per solid angle vs. total flux density per solid angle, for all sources in the high-resolution image (gray circles). Sources identified as AGN via \tbtotal\ are marked with pink stars, with sources also identified via \tbpeak\ marked with magenta stars. A line of unity is drawn to guide the eye. The high-$T_b$ sources identified by their peak brightness lie close to this line, indicating that they are unresolved. The high-$T_b$ sources identified only by their total flux density predictably have more points which lie further from this line. The right panel shows where these sources lie in terms of their total flux density measured from the standard resolution image (the x-axis), with the ratio of total flux density from the standard and high resolution images shown as the dependent variable. The dashed line at unity indicates where the total flux density in the standard resolution image is the same as measured in the high resolution image.} 
    \label{fig:obsprop}
\end{figure*}

For the flux density we make a calculation from both the peak brightness and the total flux density in the high resolution catalogue, resulting in a total of 946 unique sources selected. Of these sources, 486 were identified as AGN using the peak brightness, while all 946 were identified using the total flux density. A comparison of the peak brightness and total flux density per solid angle are shown in the left panel of Fig.~\ref{fig:obsprop}, while the right panel shows the ratio of integrated flux densities in the standard and high resolution images as a function of the integrated flux density in the standard resolution image. Sources which were identified by their peak brightness temperature appear to be largely unresolved: they lie close to the line where the peak brightness to total flux density per solid angle ratio is unity. There does not seem to be a preference for sources identified by their total flux density per solid angle to be brighter than those identified by peak brightness per solid angle. 

It is interesting to note, in the right hand panel of Fig.~\ref{fig:obsprop}, that at high flux densities the sources divide into two clear categories: some sources show a high ratio of integrated flux densities from the standard image ($S_{\mathrm{LoTSS}}$) to the integrated flux density in the high resolution image ($S_{\mathrm{ILT}}$), while others show very similar flux densities at the two resolutions. There are no sources with intermediate flux density ratios. To explore this further, we consider the sources brighter than 50 mJy (all of which are classified as AGN via their radio excess and are therefore securely AGN) and separate them into sources with high ($>7$) and low ($<7$) flux density ratios. We find that, as expected, the high ratio sample show resolved structure in both the high-resolution and standard-resolution images (except for one source that is compact in both), such that the low-resolution image encapsulates more flux. On the other hand, the low ratio sample is divided fairly evenly across three categories: those which are compact in both the high-resolution and standard-resolution images (8/31 sources), those with resolved structure in the high-resolution image but are compact or show only marginal extension in the standard-resolution image (11/31 sources), and those which show resolved structure in both images (12/31). Although the low ratio sample appear to be more compact sources in general, the high-ratio sample have smaller reported major axis values in the high-resolution image; this implies that the catalogued high-resolution size is measuring the size of just a central component (e.g. the radio core) and is not representative of the full size of the source. Given the range of properties seen, this investigation provides no clear insight as to why there are no high flux density sources with intermediate flux density ratios; future imaging of the Lockman Hole at intermediate resolutions (1\sarc\ -- 2\sarc\ ) will help answer this question.

\begin{figure}
	\centering
	\includegraphics[width=0.5\textwidth]{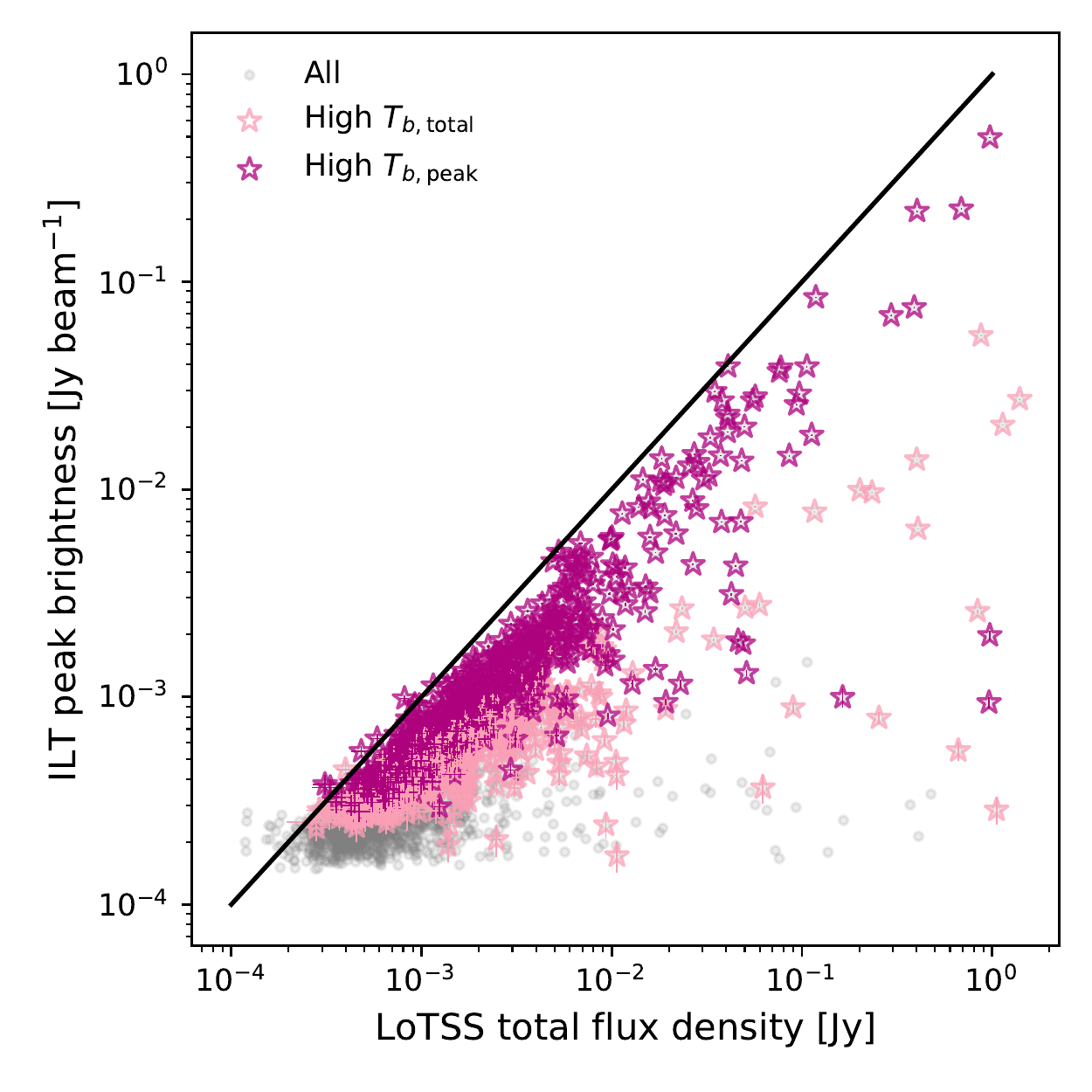}
	\caption{The peak brightness measured from the high-resolution image, compared to the integrated flux density from the standard-resolution image. The solid black line shows where these two quantities are equal. Gray circles show the entire high-resolution sample, with pink stars marking \tbtotal\ selected sources, and magenta stars marking \tbpeak\ selected sources.}
	\label{fig:fluxcomp}
\end{figure}

As a check of the quality of the catalogues, we compare the peak brightness in the high-resolution image with the integrated flux density in the standard-resolution image in Fig.~\ref{fig:fluxcomp}. We find that six of the $T_b$-identified sources have peak brightness values higher than their integrated flux density values, but overall the source population behaves as expected. The six outliers are  all at low flux densities, and four of them agree with a peak brightness to integrated flux density ratio of unity within their uncertainties. The final two agree within 3$\sigma$; they could potentially be variable sources.  Although the standard and high-resolution images were made from the same data, the standard-resolution image uses more observations, and sources in the observation selected for the high-resolution image could vary slightly from the average. We do not investigate this further here, but simply remove all six sources from the rest of our analysis, although keeping them in would not impact any of the outcomes. The final sample comprises 481 sources identified as AGN via both peak brightness and total flux density, and a further 459 identified only by total flux density (a total of 940 sources). We refer to these sources as either \tbpeak\ or \tbtotal\ in the text. Note that all sources identified as AGN via \tbpeak\ are also identified as AGN via \tbtotal ; the sources flagged as \tbtotal\ are those identified only via \tbtotal .

\begin{figure}
\includegraphics[width=0.5\textwidth]{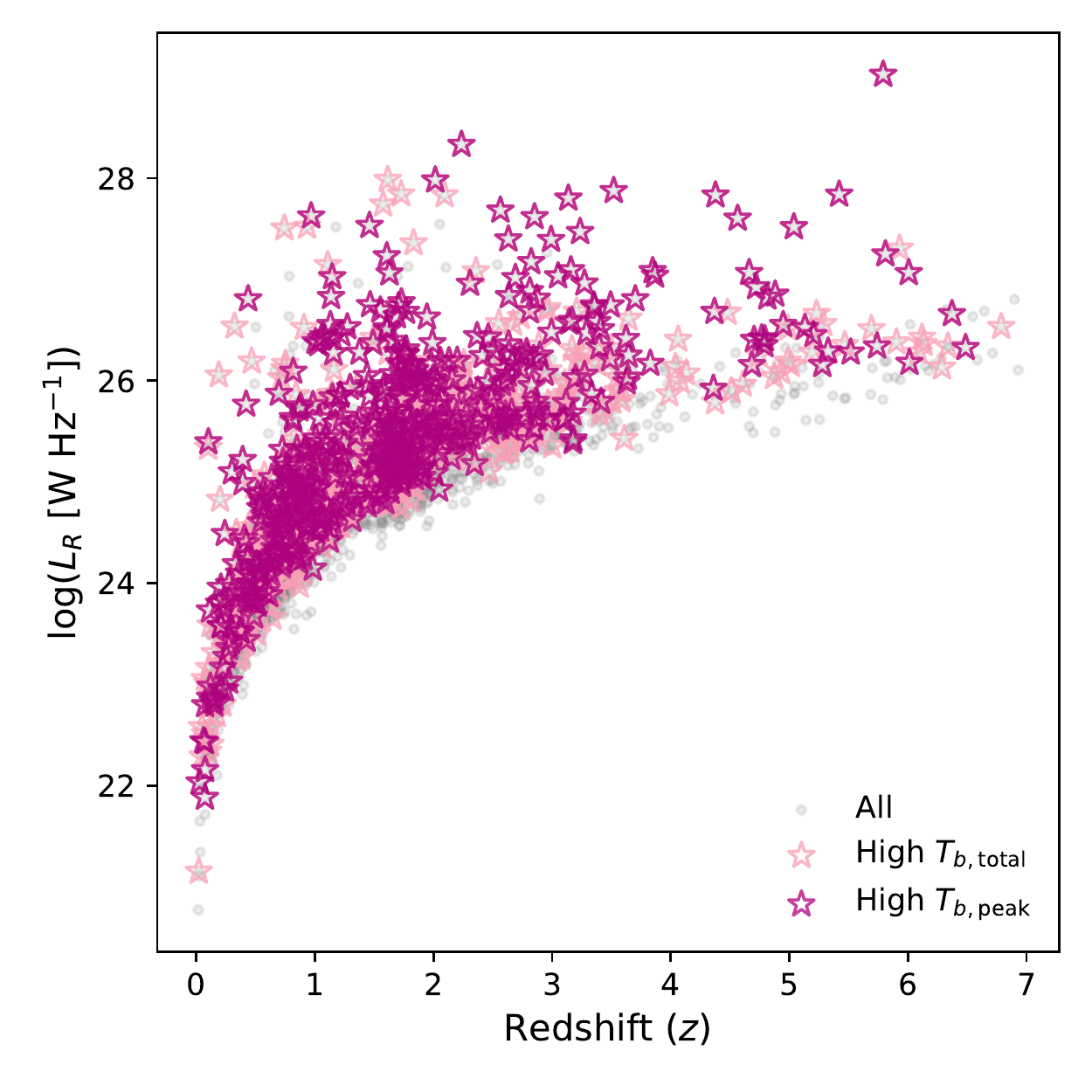}
\caption{\label{fig:pzed} Radio luminosity versus redshift for the whole sample. Error bars are not shown, because the median uncertainty on the radio power is smaller than the point size. Gray circles show the entire high-resolution sample, with pink stars marking \tbtotal\ selected sources, and magenta stars marking \tbpeak\ selected sources. }
\end{figure}

Using redshift information and assuming a spectral index of $\alpha=-0.8$, Fig.~\ref{fig:pzed} shows the integrated radio luminosity (calculated from the standard-resolution catalogue) versus redshift for both the whole high-resolution sample, and sub-samples of high-$T_b$ AGN. This shows that we are identifying AGN using their brightness temperature across all redshifts, with a wide range of radio powers. This selection method of course relies on a higher flux per beam threshold than allowed by the image sensitivity, and results in a slightly higher effective flux limit than the flux limit of the survey. 

\section{Results}
\label{sec:results}

\subsection{Source classifications}
\label{subsec:class}
Thus far we have only used the flux density and measured sizes to identify sources as AGN. We can check these identifications against those provided as part of the LOFAR Deep Fields data release, based on both photometric AGN identification \citep{kondapally_lofar_2021} and identification from detailed SED fitting (Best et al., in prep.). 

Optically identified AGN come from the Million Quasar Catalogue \citep{flesch_million_2019} compilation, IR identified AGN satisfy the \cite{donley_identifying_2012} criteria, and X-ray identified AGN have an X-ray counterpart. In Best et al., in prep. AGN are classified using two criteria, which when combined provide three AGN classes. The first criteria evaluates whether or not there is significant radiative output from an AGN; this classification is derived from the results of the 4 different SED-fitting codes, in particular through comparing the results of codes which include AGN in the fitting against those which don't. The second criteria evaluates the presence of a radio excess based on the distance from the $L_R$ -- SFR relation. Best et al., in prep. found a ridge line value for this relation of log$_{10}(L_{144}\,[$W$\,$Hz$^{-1}])=22.24+1.08\times$log$_{10}($SFR$\,[$M$_{\odot}\,$yr$^{-1}])$, and define galaxies to have a radio excess when they lie 0.7 dex (equivalent to 3$\sigma$) above this relation. The combination of these two criteria yields three classes of AGN:
\begin{itemize}
\item High excitation radio galaxies (HERGs) are radiative AGN with a radio excess
\item Low excitation radio galaxies (LERGs) are non-radiative AGN with a radio excess
\item Radio-quiet AGN (RQAGN) are radiative AGN without a radio excess
\end{itemize} 
If a classification in either category of criteria is unclear or ambiguous, the AGN identification is \textit{Unclassified}. Sources which do not fall into any of the above AGN classes are classified as star-forming galaxies (SFGs).

Table~\ref{tab:class} shows the classifications of the high-$T_b$ identified AGN\footnote{No AGN were classified via X-ray luminosity, and this criterion is not included.}, split by whether the sources were identified via \tbpeak\ and \tbtotal\ or \tbtotal\ only.  For the \tbpeak\ sample, 89 percent (427/481) also have AGN identifications from the multi-wavelength data, while this drops to 77 percent (353/459) for the \tbtotal\ sample (and is 83 percent overall considering both \tbpeak\ and \tbtotal ). This shows that our $T_b$ criterion reliably selects sources which have AGN identifications from other methods. The SED fitting may yield the wrong classification in some cases, as there is a lot of scatter around the classification lines. The classifications are also weighted by what is the bulk source of the ultraviolet through infrared emission, and in composite systems sources may be classified as an SFG when a low-luminosity AGN is present. Combining the SED and photometric AGN classifications, we also report the number of unique AGN identifications from $T_b$. There are a total of 160 new AGN identifications. 

\begin{table}
\caption{\label{tab:class} Source classification. The final column shows the number of unique AGN identifications from $T_b$ selection.}
\begin{tabular}{lcccc}
 \hline 
 \hline 
\multicolumn{5}{c}{identified with peak and total $T_b$} \\ \hline 
SED Class & \# & Opt. AGN & IR AGN & Unique $T_b$ AGN \\ \hline
SFG & 45 & 0 & 6 & 39   \\ 
Unclass & 19 & 1  & 4 & 15  \\ 
RQAGN & 31 & 11  & 21 & 0 \\ 
LERG & 292 & 0  & 8 & 0 \\ 
HERG & 94 & 30  & 44 & 0   \\ 
\textbf{Total} & \textbf{481} & \textbf{42} & \textbf{83} & \textbf{54} \\ \hline 
\hline 
\multicolumn{5}{c}{identified with total $T_b$ only} \\ \hline 
SED Class & \# & Opt. AGN & IR AGN & Unique $T_b$ AGN\\ \hline
SFG & 99 & 0 & 5 & 94   \\ 
Unclass & 14 & 1  & 2 & 12  \\ 
RQAGN & 47 & 17  & 35 & 0 \\ 
LERG & 241 & 0  & 2 & 0 \\ 
HERG & 58 & 8  & 28 & 0   \\ 
\textbf{Total} & \textbf{459} & \textbf{26} & \textbf{72} & \textbf{106} \\ \hline 
\end{tabular} 
\end{table}

\subsection{Redshift distributions}

\begin{figure}
    \centering
    \includegraphics[width=0.5\textwidth]{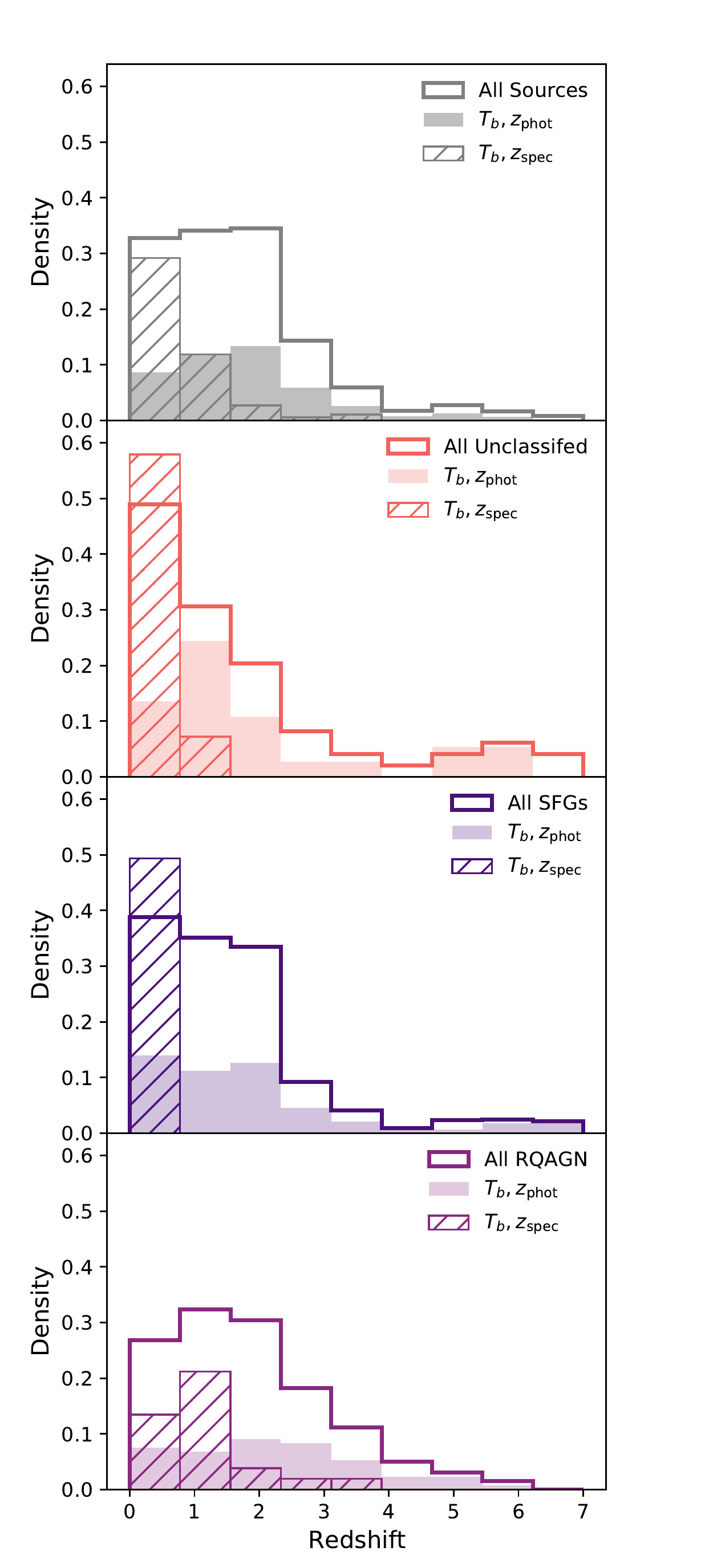}
    \caption{The overall redshift distribution for all sources in a class (all, unclassified, SFG, or RQAGN, from top to bottom panel) is plotted as a solid line. The hatched and shaded regions show the spectroscopic and photometric redshift distributions, respectively, for sources which are identified as AGN via $T_b$. The $T_b$ distributions are normalised to the overall density of all sources in each subclass: e.g., if the hatched and shaded regions are added together the area under the resulting distribution would be unity.}
    \label{fig:zinfo}
\end{figure}

Only 191 ($\sim$9 percent) of the sources have spectroscopic redshifts. If photometric redshifts are inaccurate then the limiting $T_b$ value for AGN identification can also be inaccurate, although using the redshift rather than assuming $z=0$ when calculating the limiting value of $T_b$ only adds 27 sources (3 percent) to the sample, so this is not expected to be a large effect. \cite{duncan_lofar_2021} shows that the photometric redshifts are reasonably accurate out to $z\sim 4$, with $\sigma_{\textsc{NMAD}}=0.017$ and $0.077$ for galaxy/host-dominated and AGN, respectively. Fig.~\ref{fig:zinfo} shows that some of the Unclassified, SFG, and RQAGN sources have primary redshift solutions beyond this, but out of the 2214 sources which had SED fitting performed this is a small number, only 112 ($\sim$5 percent). What is clear is that the Unclassified and SFG sources may be mis-identified or not identified because their redshift is insecure; there is a dearth of $z_{\mathrm{spec}}>0.5$ sources in these classes, which is not true for the RQAGN class. The long tails of photometric redshifts for the Unclassified and SFG categories may change if followed up to acquire spectroscopic redshifts, which is an interesting prospect for a future study. 
lie

\subsection{Infrared colour-colour plots}

\begin{figure}
    \centering
    \includegraphics[width=0.45\textwidth]{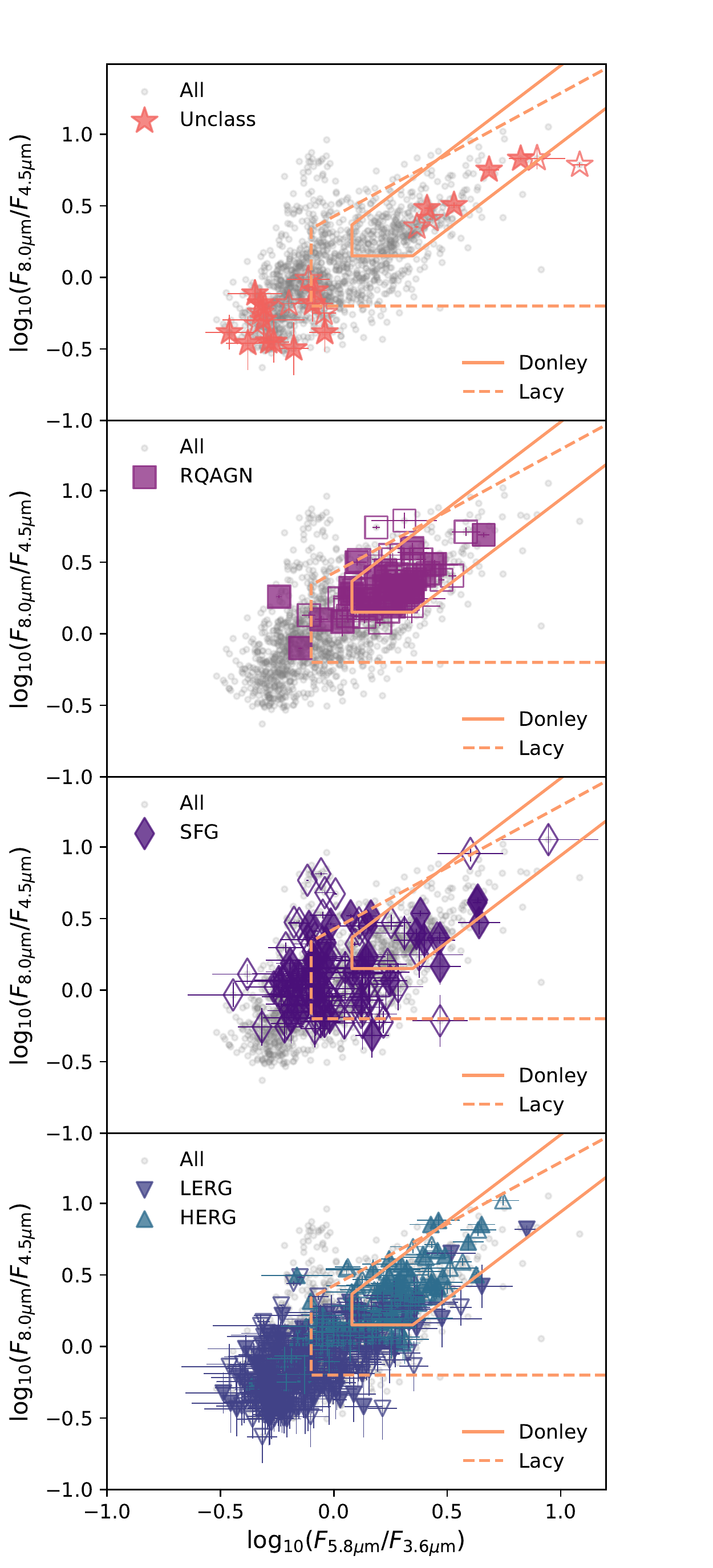}
    \caption{Infrared colour-colour plots. The gray points show all sources detected in the high-resolution image, while the coloured points show the $T_b$-identified AGN, using the SED-fitting classifications. Filled points are identified via $T_{b,peak}$ while unfilled points are identified only via $T_{b,total}$. Note that only sources with $>$2$\sigma$ detections in all four bands are plotted. Wedges which are often used to identify AGN from their infrared colours are shown in orange. }
    \label{fig:donley}
\end{figure}

We construct the infrared colour-colour plots using the multi-wavelength catalogue from \cite{kondapally_lofar_2021}, for sources with $>2\sigma$ detections in all four IRAC channels (3.6, 4.5, 5.8, and 8.0$\,\mu$m). Figure~\ref{fig:donley} shows where they lie in relation to the wedges from \cite{donley_identifying_2012} and \cite{lacy_obscured_2004} over-plotted. Unsurprisingly, given that the mid-IR region of the SED is an important driver of the classifications of Best et al (in prep), the RQAGN fall mostly in the wedges which identify them as radiative AGN, as do the HERGs. The LERG population extends down to the lower left of the colour-colour plot. The bulk of the SFG population seems to hover around the edge of the wedges, suggesting that perhaps they are composite SFG/AGN sources. The Unclassified category splits into two groups: those which fall securely in the AGN wedges, and those which lie below the wedges. It is worth keeping in mind that the position of sources on these plots will vary with redshift \citep{radcliffe_nowhere_2021}.  

\subsection{The $L_R$ -- SFR relation}

\begin{figure}
    \centering
    \includegraphics[width=0.5\textwidth]{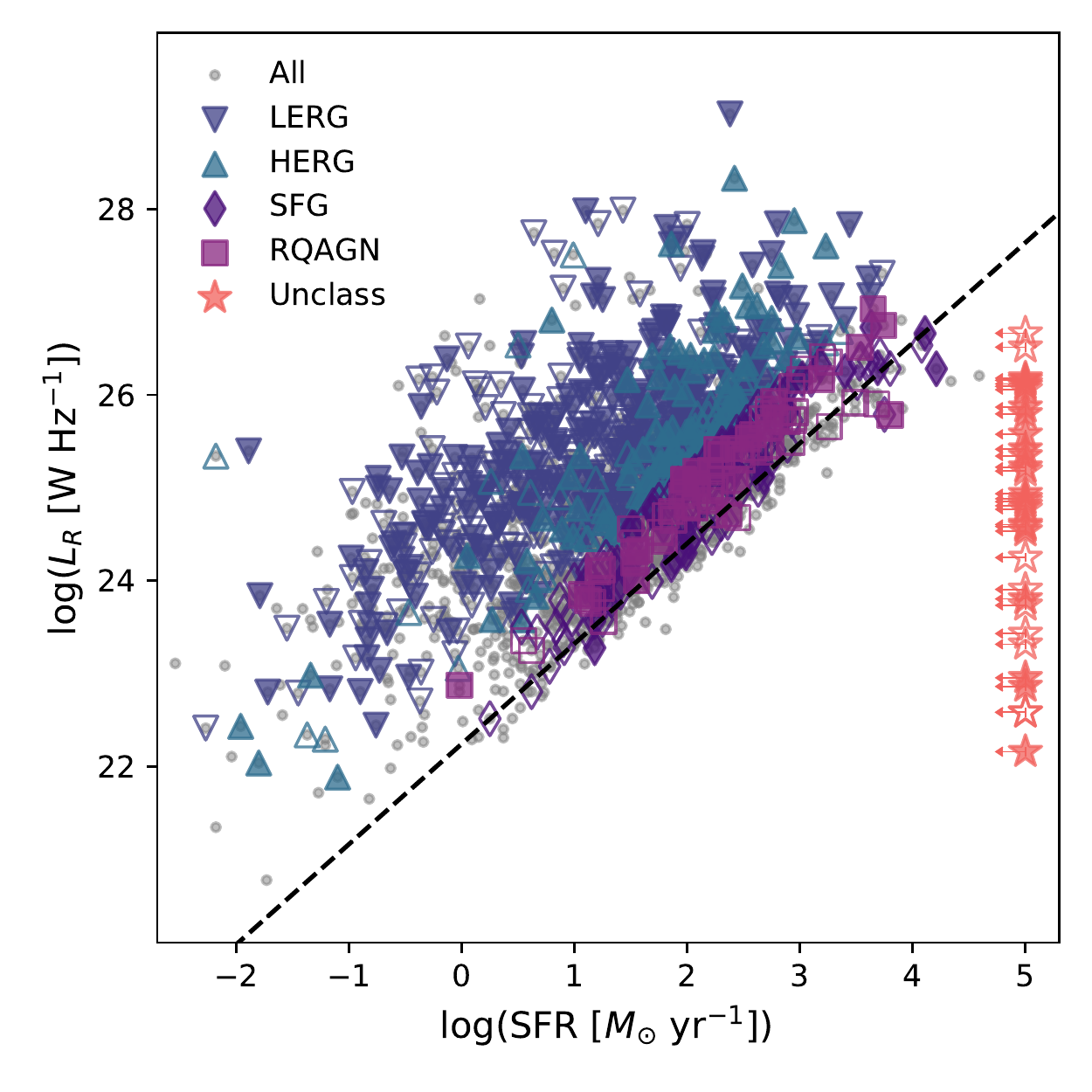}
    \caption{The $L_R\,\sim\,$SFR relation. The line above which sources are classified as radio excess is not explicitly drawn, but clear from the lower limit of where LERGs/HERGs fall on the plot. $T_b$-identified AGN run the full range of star formation rates and luminosities. The unclassified sources do not have estimated SFR values, and are shown on the right hand side of the plot to illustrate their range of $L_R$. The ridge line from Best et al. (in prep) described in \S~\ref{sec:sfagn} is shown as a dashed black line.}
    \label{fig:lrsfr}
\end{figure}

We can also look at the source classifications in terms of the $L_R$ -- SFR relation, in Fig.~\ref{fig:lrsfr}. Although the sources unclassified by the SED fitting do not have estimated SFR values, we place them on the plot to the right hand side to show their $L_R$ values. It is clear that in general the high-$T_b$ identified sources run the range of SFR and $L_R$ values. For non-radio excess sources, there is avoidance of the ridge line from Best et al., in prep.: 87 percent of the high-$T_b$ SFGs (125/144) and 91 percent of the high-$T_b$ RQAGN (71/78) are above the ridge line of the $L_R$ -- SFR relation (but, by definition, not 0.7 dex above it, which would classify them as radio-excess sources). This may suggest the presence of AGN-related radio emission, even if not formally defined as radio excess sources.

\subsection{Detectability and source fractions}

\begin{figure}
	\centering
	\includegraphics[width=0.5\textwidth]{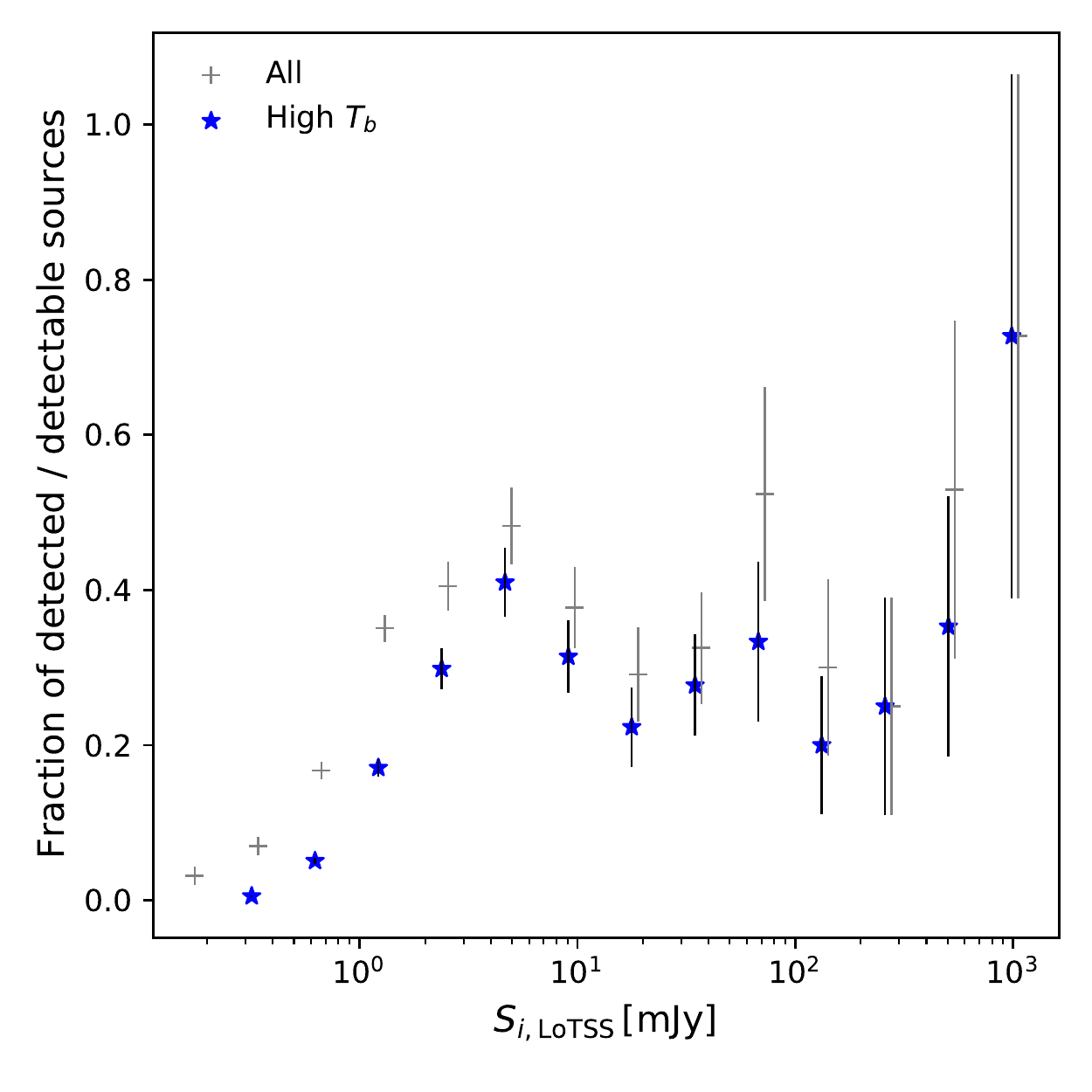}
	\caption{Fraction of detected sources (of those which are detectable) as a function of $S_{i,\mathrm{LoTSS}}$. The gray symbols show this fraction for all sources, while the blue stars show this for the $T_b$-selected sources. The distribution increases up to $\sim$5 mJy, then remains roughly constant (within uncertainties) at higher flux densities. This could correspond to the switch between the SFG and AGN dominated populations. Uncertainties are estimated assuming $\sqrt{N}$ for $N$ sources in a bin, and propagated in the standard manner. }
	\label{fig:detfrac}
\end{figure}

In the standard-resolution image, there are 23,734 sources with SNR$>5$ which are within the coverage of the high-resolution image. Not all of these will be detectable at high resolution, as they may be extended low-surface brightness sources. To determine the number of sources which are detectable, we compare the peak brightness from the standard-resolution image to the local rms in the high-resolution image. There are 13,439 sources for which the peak brightness is larger than 5 times the rms noise at that location in the rms map, which are potentially detectable. Only 16 percent (2214) of these sources are detected in the high-resolution image, which is similar to what is found in studies carried out at higher radio frequencies.  High-resolution radio follow up with VLBI arrays of deep radio samples have shown a detection rate of 20 -- 30 percent, depending on the sample \cite[e.g.][]{herrera_ruiz_faint_2017,radcliffe_nowhere_2018}. Figure~\ref{fig:detfrac} shows the fraction of detected to detectable sources as a function of integrated flux density from the standard-resolution image. The high-resolution sample is of course biased towards compact sources, as sources with low surface brightness are less likely to be detected in the high-resolution image. While the high-$T_b$ sources (we remind the reader that this includes all \tbpeak\ and \tbtotal\ sources) do not differ from the overall population of detected sources (within the uncertainties) above $\sim$5$\,$mJy, there is a clear separation of the high-$T_b$ sources below this. Sources with higher flux densities are more likely to be radio-loud, and therefore AGN dominated, whereas we expect star formation to dominate at fainter flux densities \citep[e.g.,][]{ibar_deep_2009,ocran_nature_2017,prandoni_lockman_2018}. 

\begin{figure*}
	\centering
	\includegraphics[width=0.45\textwidth]{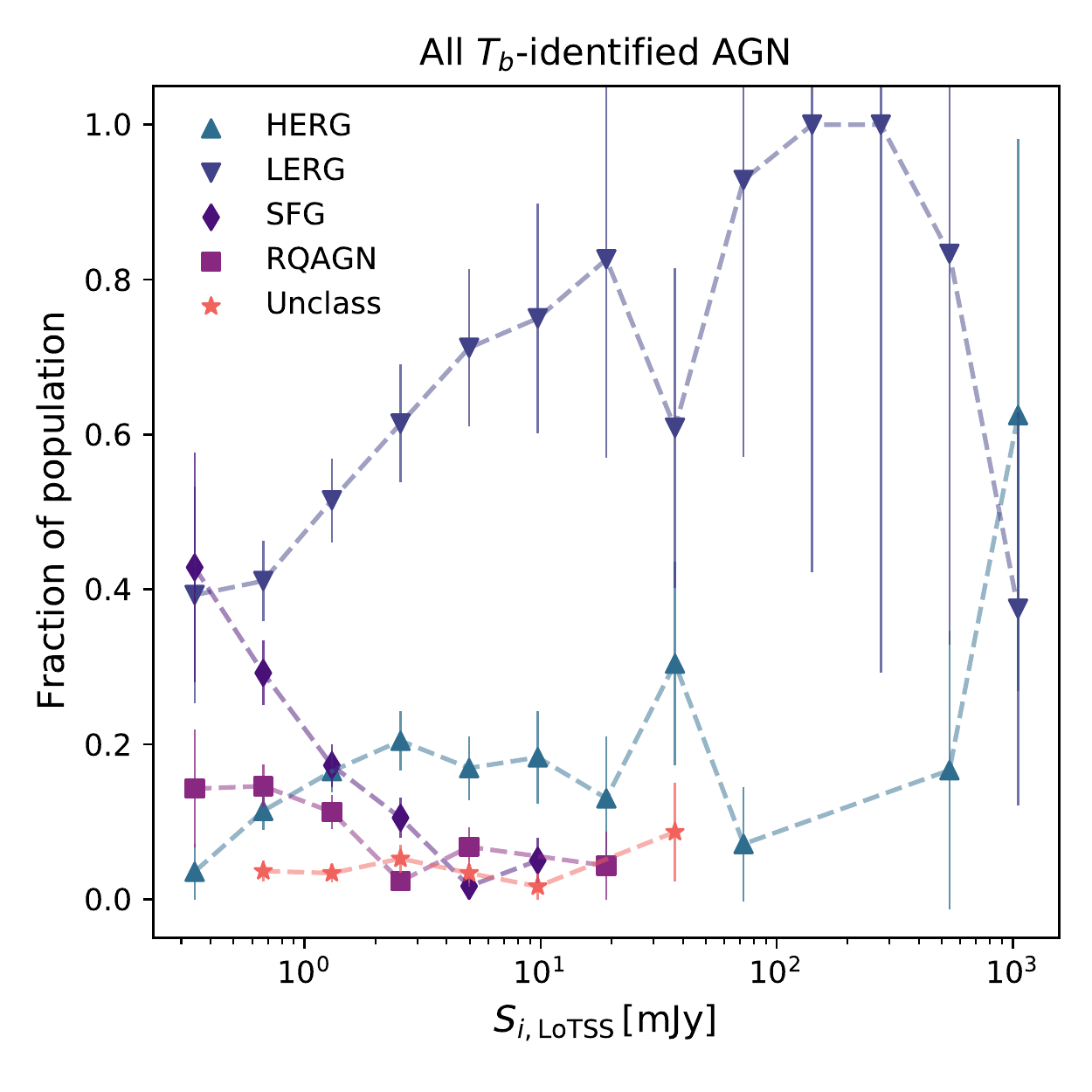}
		\includegraphics[width=0.45\textwidth]{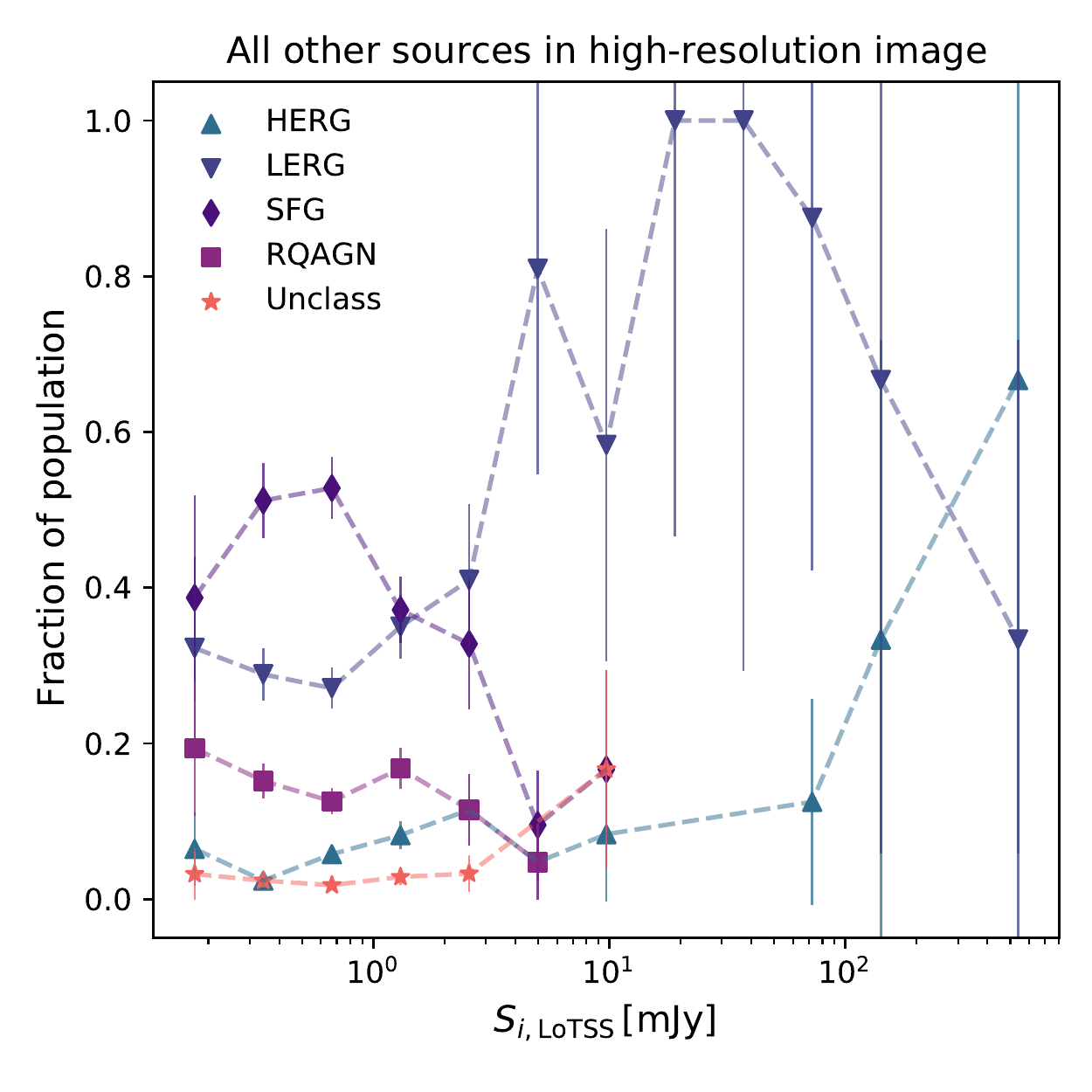}
	\caption{Here we plot the composition of the $T_b$-identified AGN (left panel) and all other sources (right panel) as a function of integrated flux density measured from the standard-resolution image. In each flux density bin, the fraction of each sub-population of the total number of sources in the bin is plotted.}
	\label{fig:sourcefracs}
\end{figure*}

Finally, we investigate the nature of the high-$T_b$ sample, and compare it to the non-$T_b$ selected sources in the high-resolution image (for comparison with the full sample detected in the standard-resolution image, see Fig. 9 of Best et al. in prep). For each sample, $T_b$-identified AGN and non-$T_b$ selected sources, we split them into their sub-populations as a function of flux density. As above, we use the total flux density measured from the standard-resolution image. 
We calculate the number of high-$T_b$ sources in each flux density bin, and find the fraction of each sub-population. 
The results are shown in the left ($T_b$-identified AGN) and right (non-$T_b$ selected sources) panels of Fig.~\ref{fig:sourcefracs}. It is clear that the LERGs dominate the both samples, particularly at higher flux densities, with the HERG contribution increasing at the highest flux densities. This is the same behaviour seen in Best et al., in prep. (their Fig. 9).  In the $T_b$-identified AGN sample, the SFG population is different: in Best et al., in prep. this dominates below $\sim$1$\,$mJy, whereas here the SFG population fraction remains low for most of the flux density range, only rising to 0.4 in the lowest flux density bin. This is because the SFGs should all be excluded from the current sample by the high-$T_b$ identification. We see that the SFGs are contributing the most to the non-$T_b$ selected sources below flux densities of $\sim$2 mJy, which is consistent with Best et al. in prep. 

\begin{figure}
	\centering
		\includegraphics[width=0.45\textwidth]{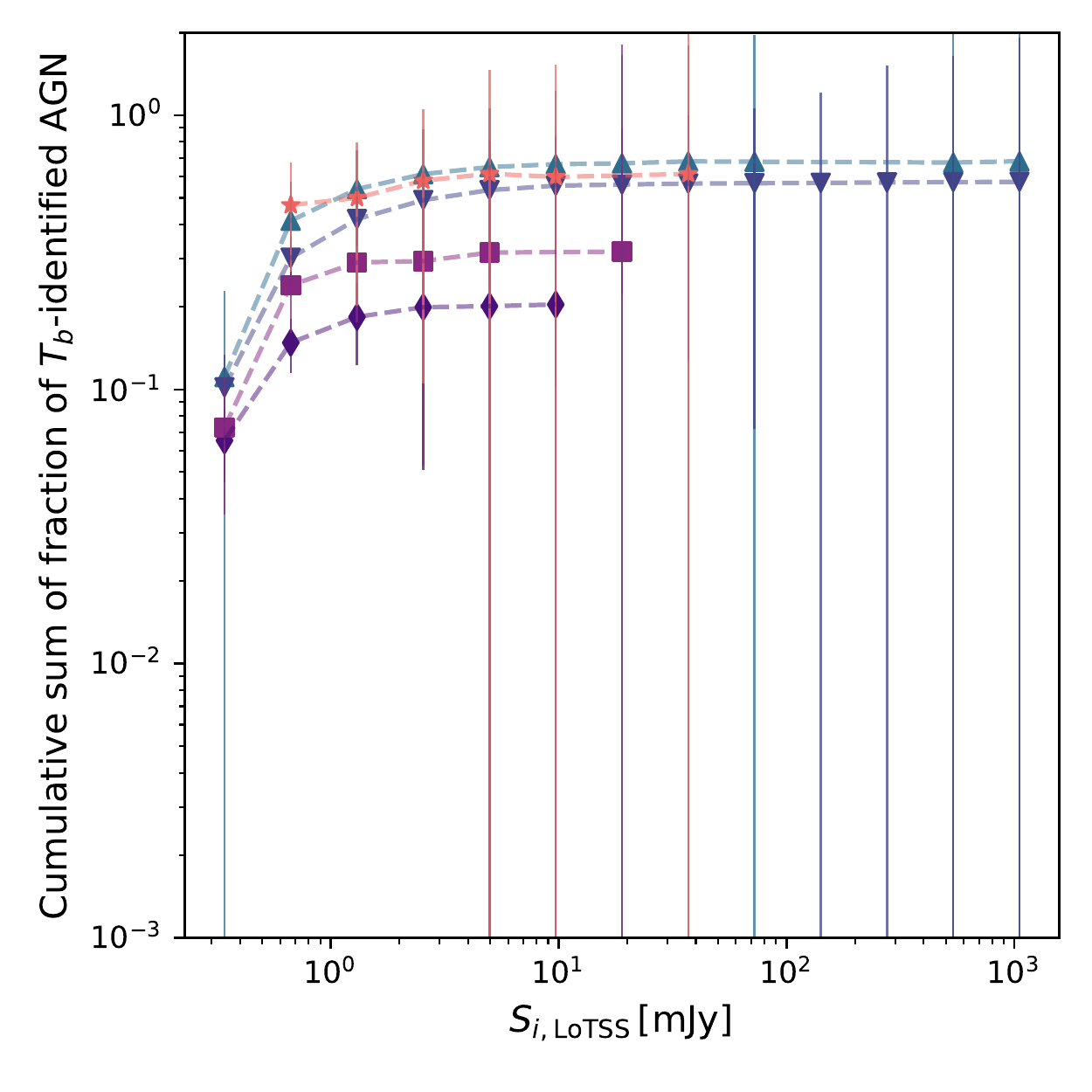}
		\caption{The cumulative sum of the fraction of $T_b$-identified AGN in each sub-population, as a function of integrated flux density as measured from the standard-resolution image. Symbols and lines are the same as in Fig.~\ref{fig:sourcefracs}. The lines flatten out at the total fraction of each sub-population which are $T_b$-identified AGN. }
	\label{fig:fracpop}
\end{figure}

Figure~\ref{fig:fracpop} shows the cumulative sum of the fraction of high-$T_b$ sources in the respective sub-population. The approximate completeness of the $T_b$ identification is apparent: the cumulative sum of the fractions asymptote above $\sim$1 mJy. The overall fraction of high-$T_b$ sources in each overall sub-populations is: HERGs -- 68 percent; LERGs -- 57 percent; Unclassified -- 61 percent; RQAGN -- 32 percent; SFG -- 20 percent. The HERG, LERG, and unclassified populations have the largest fractions of high-$T_b$ sources, which is a further indication that the unclassified sources are indeed AGN. The RQAGN and SFG populations follow with smaller fractions of their sub-populations comprising high-$T_b$ sources. For the RQAGN, this implies that the population may not be dominated by one type of radio emission mechanism, but instead may comprise sources with different types of radio emission \citep[e.g., jets, winds, star formation; see][]{panessa_origin_2019}.  It is also possible that some sources identified as SFGs are actually RQAGN, which would increase the number of $T_b$-identified RQAGN here (although this number is expected to be a lower limit, see \S~\ref{subsec:tb}), leaving a very small fraction of genuine SFGs potentially hosting radio jets. Either way, it seems clear that high-$T_b$ AGN identifications are preferentially associated with radiative AGN (RQAGN) than SFGs.

\section{Separating star formation and AGN activity}
\label{sec:sfagn}

We turn to separating radio emission from star formation and AGN activity, which relies on understanding how star formation contributes to the brightness temperature, as described in Section~\ref{sec:tb}. We do this only for $T_b$-identified sources which are unresolved in the high-resolution image and do not have a radio excess, i.e. the SFG, RQAGN, and Unclass populations. This avoids sources which may have extended or resolved emission due to AGN activity. We determine the unresolved population in the same way as \cite{shimwell_lofar_2019,shimwell_lofar_2022}, by fitting an upper envelope at the 99.9th percentile to the sources fit with a single Gaussian component (code {\tt `S'}). This was done using the high-resolution image. A total of 53 $T_b$-identified sources (31 \tbpeak , 22 \tbtotal ) were determined to be resolved and therefore removed.

We make the assumption that star formation is widespread in a galaxy, producing a low surface brightness component that will not be detected in the high-resolution image, while the AGN component, identified via brightness temperature, will be a single compact element. Thus the peak brightness in the high-resolution image will represent the AGN luminosity, although this may be underestimated because of smearing (see \S~\ref{subsec:tb}). We use the peak brightness and not the total flux density to avoid contamination from potential low-surface brightness contributions (i.e., star formation). 
We make use of the standard-resolution image to estimate the contribution from star formation. Unless very nearby, galaxies will be unresolved in the standard-resolution image, and low-surface brightness radio emission from star formation will be above the limit for detection. 
Forward-modelling the problem using 3D models of galaxies that are `observed' in the same way as these data will, in a future paper, allow us to place more secure constraints on the separation of star formation and AGN activity. 

With these assumptions in hand, we calculate the following AGN and star formation contributions:
\begin{equation}
    L_{\mathrm{AGN}} = L_{p,\mathrm{ILT}} 
\end{equation}
\vspace{-12pt}
\begin{equation}
    L_{\mathrm{SF}} = L_{i,\mathrm{LoTSS}} - L_{p,\mathrm{ILT}}
\end{equation}
where $L_{p,\mathrm{ILT}}$ is the luminosity obtained using the peak intensity in the high-resolution image, and $L_{i,\mathrm{LoTSS}}$ is the luminosity obtained from the integrated flux density in the standard-resolution image. Errors are propagated in the standard way. 
The SFR contribution to the radio luminosity can be converted into SFR using the best fit parameters for Eqn. 2 of \cite{smith_lofar_2021}:
\begin{equation}
    0.9 \times \mathrm{log}_{10}(\textrm{SFR}) = \mathrm{log}_{10}(L_{\mathrm{SFR}}) - 22.22 - 0.33 \times \mathrm{log}_{10}(M_{*}) 
\end{equation}
where the units of SFR and $M_*$ are $M_{\odot}\,\mathrm{yr}^{-1}$ and $10^{10}M_{\odot}$, respectively. This equation assumes that there is no low frequency absorption. This mass-dependent form is similar to the Best et al. (in prep) ridge line. Practically, we calculate $S_{\mathrm{AGN}}$ and $S_{\mathrm{SF}}$ and convert the flux densities into rest frame luminosities using the distance modulus, assuming a typical synchrotron spectrum of $\alpha=-0.8$. 

\begin{figure}
    \centering
    \includegraphics[width=0.5\textwidth]{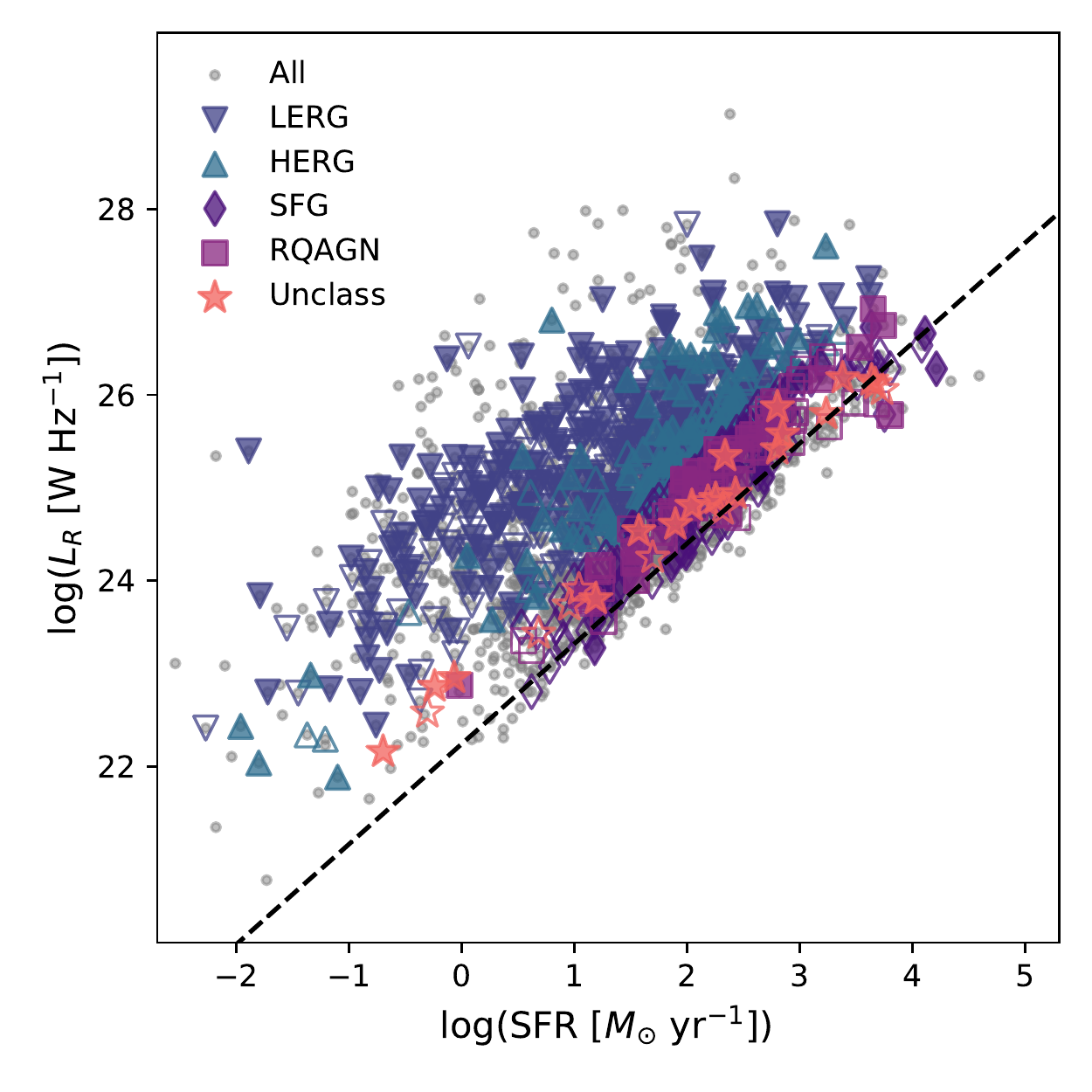}
    \caption{Same as Fig.~\ref{fig:lrsfr} for all sources but the unclassified category. For these, we have only been able to place them on the plot after estimating their star formation rate by subtracting the $L_{\mathrm{AGN}}$ contribution to the total radio luminosity.}
    \label{fig:radiosfrclass}
\end{figure}

Using the total radio luminosity and the measured SFR after subtracting the AGN component, we can now place our unclassified sources in the $L_R$ -- SFR parameter space, see Fig.~\ref{fig:radiosfrclass}. None of the unclassified sources shows a radio excess based on the SFR, although they lie on the upper half of the $L_R$ -- SFR relation, with 73 percent (22/30) of sources lying above the ridge line.  

\begin{figure*}
    \centering
    \includegraphics[width=1\textwidth]{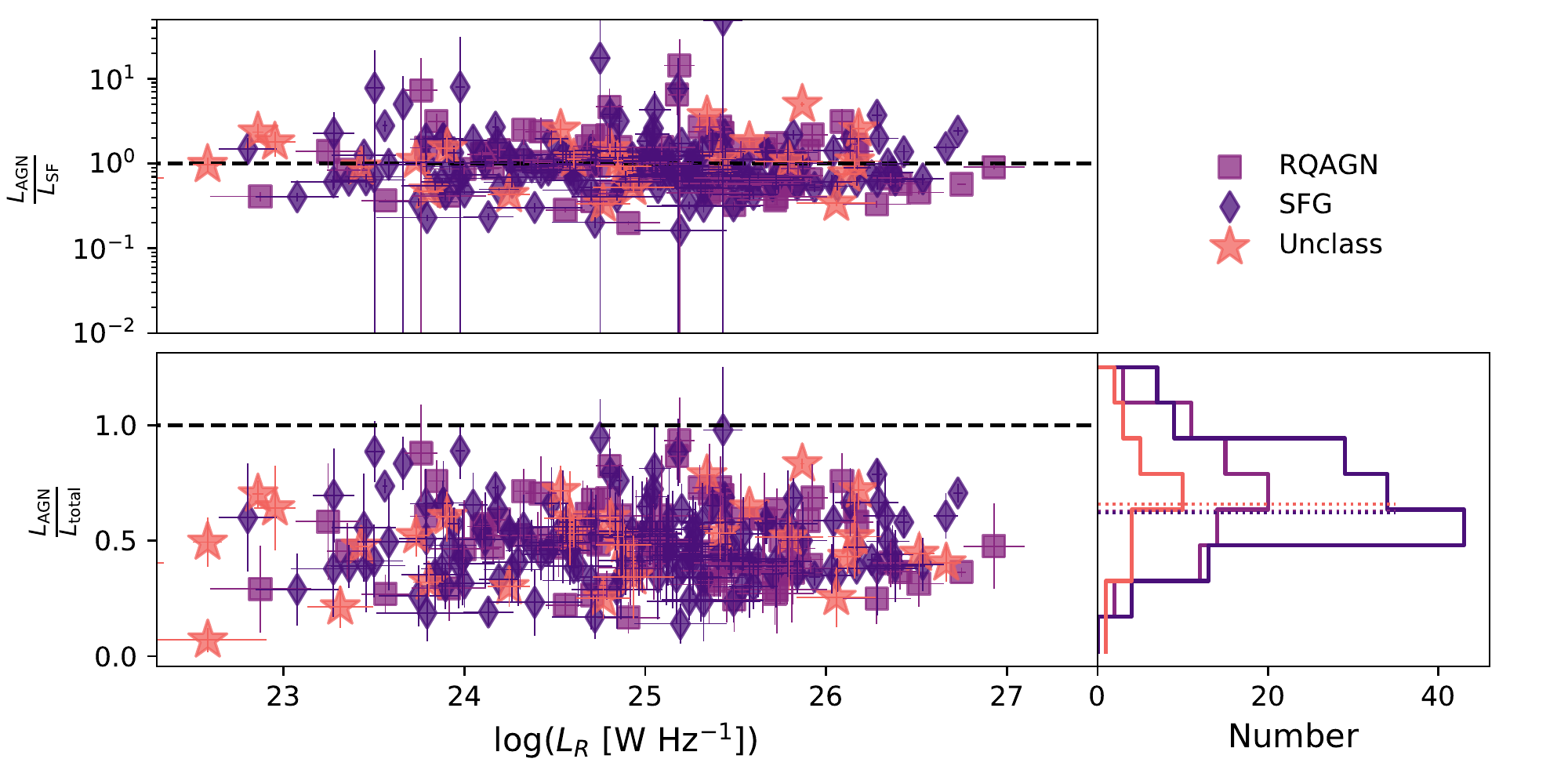}
    \caption{The top panel shows the ratio of radio luminosity from AGN to radio luminosity from star formation, as a function of total radio luminosity. The bottom panel shows the ratio of AGN to total radio luminosity, with the distributions of the AGN fraction for each population in the panel to the right (dotted lines show the medians reported in Table~\ref{tab:fracs}). The black dashed line in each main panel is placed at unity to help guide the eye. }
    \label{fig:agnfracs}
\end{figure*}

We investigate the relative contributions of the AGN to the total radio luminosity, in Fig.~\ref{fig:agnfracs}. The top panel of this figure shows the ratio of $L_{\mathrm{AGN}}/L_{\mathrm{SF}}$, while the bottom panel shows $L_{\mathrm{AGN}}/L_{\mathrm{total}}$, both as a function of total radio luminosity. The three populations are well mixed, with no one population having significantly higher or lower fractions than the others. We report the median values in Tab.~\ref{tab:fracs}. We use the median (with the median absolute deviation for the uncertainties) rather than the mean as it is more robust against outliers. The values generally agree with each other: when $L_{\mathrm{AGN}}/L_{\mathrm{total}}$ is $\sim$0.5, $L_{\mathrm{AGN}}/L_{\mathrm{SF}}\sim 1$, as expected. Although the ratio of $L_{\mathrm{AGN}}/L_{\mathrm{total}}$ is slightly lower in the SFG category, all three types of sources have medians which agree within the uncertainties. This is consistent with what we would expect: an ILT detection means there must be a compact component, and high values of $T_b$ mean there is an AGN component, so $L_{\mathrm{AGN}}/L_{\mathrm{SF}}$ will always be above zero. Additionally, we have limited this part of the analysis to sources which do not have a radio excess, so the total emission will always be less than $\sim$5 times that expected from SFR, meaning that the AGN will not account for more than $\sim$80 percent of the total radio emission. 

\begin{table}
 \centering 
\caption{\label{tab:fracs} Median values of ratios from Fig.~\ref{fig:agnfracs}.}
\begin{tabular}{lc}
 \hline\multicolumn{2}{c}{$L_{\mathrm{AGN}}/L_{\mathrm{SF}}$} \\ \hline 
Class & Median\\ \hline
RQAGN & 0.98$\pm$0.67 \\ 
SFG & 0.96$\pm$0.55 \\ 
Unclass &  1.07$\pm$0.68 \\ 
\textbf{Total} & \textbf{ 0.98$\pm$0.60 } \\ \hline \hline  
\multicolumn{2}{c}{$L_{\mathrm{AGN}}/L_{\mathrm{total}}$} \\ \hline 
Class &  Median\\ \hline
RQAGN & 0.49$\pm$0.17 \\ 
SFG & 0.49$\pm$0.15 \\ 
Unclass & 0.52$\pm$0.17 \\ 
\textbf{Total} & \textbf{ 0.49$\pm$0.16 } \\ \hline \hline  
\end{tabular} 
\end{table}

We also fit a linear model to $L_{\mathrm{AGN}}/L_{\mathrm{total}}=m \times $log$_{10}(L_R) + b$ to check for any dependence between radio and AGN luminosity. The fit results are reported in Tab.~\ref{tab:fracfits}, which shows at most only very mild dependence, with relatively large uncertainties. This may be impacted by the effective flux limit of the $T_b$ selection (see Fig.~\ref{fig:pzed}), imperfect separation between SFR and AGN radio emission, or due to the fact that we are probing a regime where neither AGN nor SFR radio luminosity dominates the total radio luminosity. Future work to more precisely quantify the uncertainties on the AGN luminosity (using forward modelling) will help us understand the picture here. 

\begin{table}
 \centering 
\caption{\label{tab:fracfits} Fit parameters for $L_{\textrm{AGN}}/L_{\mathrm{total}}=m \times $log$_{10}(L_R) + b$ from Fig.~\ref{fig:agnfracs}.}
\begin{tabular}{lcc}
 \hline 
Class & $m$ & $b$ \\ \hline
All & $-0.00\pm$0.01  & $0.52\pm$0.29 \\ 
RQAGN & $-0.03\pm$0.02  & $1.15\pm$0.58 \\ 
SFG & $-0.00\pm$0.02 & $0.57\pm$0.40 \\ 
Unclass &  $0.03\pm$0.03  & $-0.22\pm$0.63 \\ \hline 
\end{tabular} 
\end{table}

\begin{figure}
	\centering
	\includegraphics[width=0.5\textwidth]{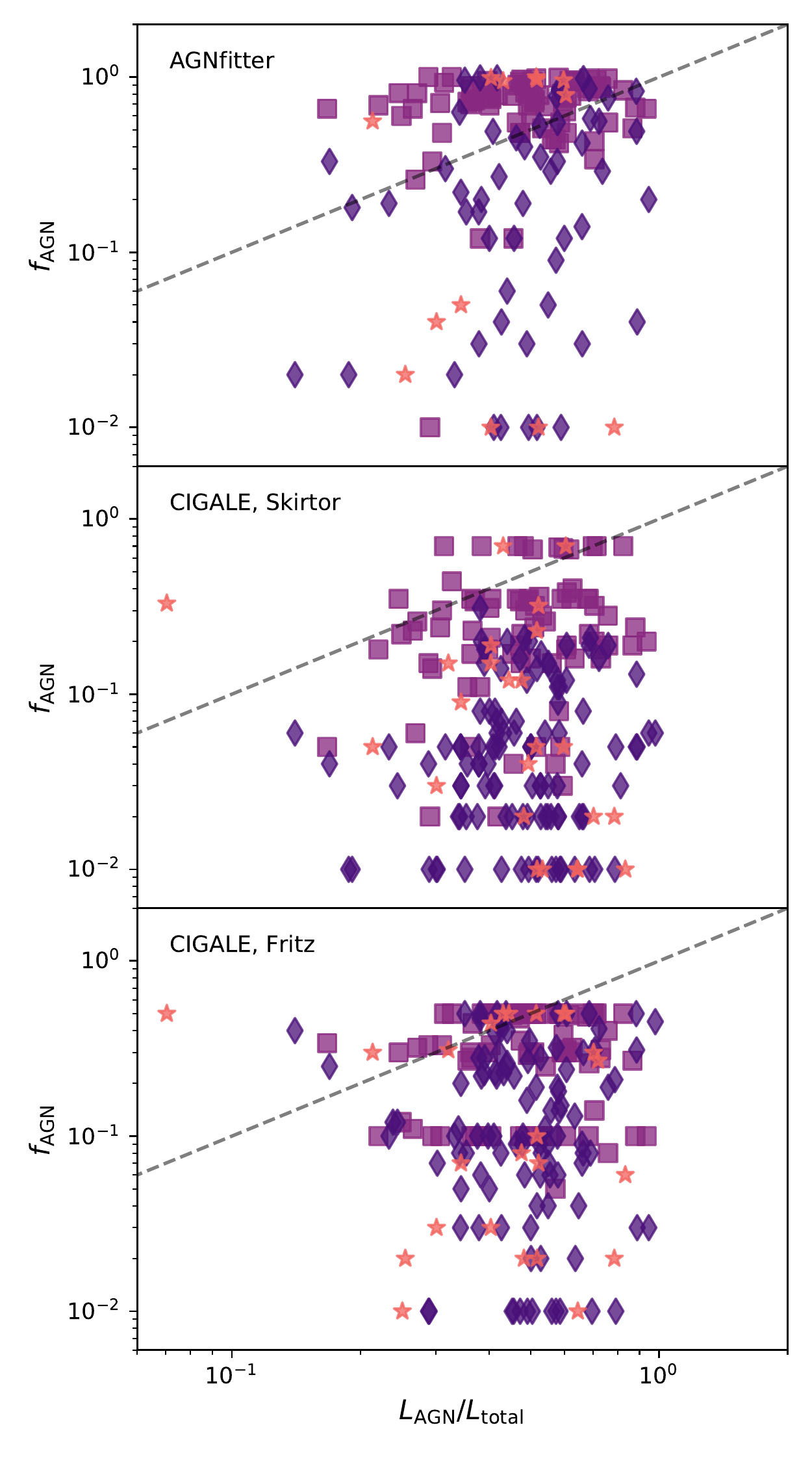}
	\caption{Comparison of the measured $L_{\mathrm{AGN}}/L_{\mathrm{total}}$ with the SED-estimated $f_{\mathrm{AGN}}$  values.  The dashed line in each plot shows where $L_{\mathrm{AGN}}/L_{\mathrm{total}} = f_{\mathrm{AGN}}$. Note that only sources which have a value of $f_{\mathrm{AGN}}$ reported are shown. Symbols are the same as in Fig.~\ref{fig:agnfracs}. }
	\label{fig:fracs}
\end{figure}

The SED fitting also calculates an AGN fraction, which we compare with $L_{\mathrm{AGN}}/L_{\mathrm{total}}$ in Fig.~\ref{fig:fracs}. As there are multiple SED fits to find consensus values, we show the $f_{\mathrm{AGN}}$ for \textsc{AGNfitter} and \textsc{CIGALE}, the latter with both the \textsc{SKIRTOR} \citep{stalevski_3d_2012,stalevski_dust_2016} and \cite{fritz_revisiting_2006} AGN models. It is clear that there is a wide range of $f_{\mathrm{AGN}}$ values regardless of which SED fitting method is used. This is partly dependent on the fact that $f_{\mathrm{AGN}}$ is defined differently for each SED fitting method. For \textsc{AGNfitter} it is the fraction of emission in the 1-30 micron range from the AGN components. For \textsc{CIGALE} it is the total IR luminosity fraction, which naturally will give lower values than \textsc{AGNfitter}.  For a LERG-type source, where the dominant AGN output is in the radio band, SED fitting can yield low values of $f_{\mathrm{AGN}}$; this is consistent with at least some of the SFGs and Unclassified sources not being identified as LERGs because they are not above the radio excess definition, while the majority of RQAGN still have high $f_{\mathrm{AGN}}$.  It is clear that we do not (except for an outlier) measure $L_{\mathrm{AGN}}/L_{\mathrm{total}} \lesssim 0.2$, while the SED fitting measures values of $f_{\mathrm{AGN}}$ as small as 0.01; as $L_{\mathrm{AGN}}/L_{\mathrm{total}}$  sample is biased towards AGN selected via $T_b$ in the radio band, this is not surprising.  Aside from some outliers, the RQAGN population seems to show better agreement between $f_{\mathrm{AGN}}$ and $L_{\mathrm{AGN}}/L_{\mathrm{total}}$ for all three SED fitting methods, which is not totally unexpected as they have been identified as AGN via SED fitting. For \textsc{AGNfitter} and \textsc{CIGALE} with the \textsc{Fritz} models, there seem to be a population of SFGs with higher relative $f_{\mathrm{AGN}}$ values that correlate with $L_{\mathrm{AGN}}/L_{\mathrm{total}}$, although not with a one-to-one relation. Overall, the measured $L_{\mathrm{AGN}}/L_{\mathrm{total}}$ value is higher than that estimated by the SED fitting code: we are either over-estimating the AGN contribution or the SED fitting is under-estimating the AGN contribution -- however, this must be tempered by the fact that $f_{\mathrm{AGN}}$ and $L_{\mathrm{AGN}}/L_{\mathrm{total}}$ are measured differently and may not be like comparisons. The AGN contribution to the radio emission may also be less direct in these non-radio excess sources (e.g., if it comes from secondary effects like shocks from AGN winds).

\section{Discussion}
\label{sec:discussion}

\subsection{Caveats}
\label{subsec:caveats}

Identification of AGN using brightness temperature is a widely used method, but this is the first time it has been applied at low observing frequencies. One of the primary things to keep in mind is that the models to calculate the limiting value of brightness temperature (practically, here we use flux density per solid angle) assume only the simplest model of a free-free absorbing medium  that is well mixed with the radio synchrotron emitting plasma, and in which there is a constant ratio of these two components.  Other low frequency cut-off mechanisms can exist such as synchrotron self-absorption and low energy electron cut-off.  Additionally the geometry of any free-free absorbing gas could be different from the assumptions of the Condon model, in particular it could be clumpy by being confined to \textsc{H}$\,$\textsc{ii} regions \citep{lacki_interpreting_2013,ramirez-olivencia_sub-arcsecond_2022} or could be foreground to the synchrotron emission \citep{varenius_subarcsecond_2015,conway_continuum_2018}. Such different geometries affect the predicted radio spectrum from a star-formation powered source and hence the maximum brightness temperature expected at a given frequency. 

In addition to absorption with a covering factor less than unity across the whole source, AGN activity and star formation in galaxies which host both can experience different types of absorption in each. For example, if the compact AGN emission comes from jets which are experiencing synchrotron self-absorption, radio emission from star formation in the galaxy would be unaffected. In the case of free-free absorption of radio emission from star formation due to ionising gas along the line of sight, whether or not it impacts the compact AGN emission depends on the geometry of the system and the covering factor of the ionising gas. Of course, the covering factor could be unity and provide a constant amount of absorption across the entire source, which means the \textit{relative} contributions, if not the \textit{absolute}, of star formation and AGN activity would remain the same. Regardless of its origin, low-frequency absorption would work to suppress the flux density per solid angle. This means that while the our conservative limit ensures we are securely identifying AGN, many other ILT sources may also be AGN. 

Another issue to consider is the beam size. In high-frequency observations, the spatial scales probed are smaller. For example, \cite{middelberg_mosaiced_2013} achieves a beam size (with natural weighting) of $11.7\times 9.4\,$mas$^{2}$, which is a beam solid angle 11 times smaller than that achieved with the ILT in \cite{sweijen_deep_2022}, $30\times 40\,$mas$^2$ (in principle, this could be pushed to $\sim$20 mas, at the cost of increased noise). So are we actually probing the relevant spatial scales? We are secure in our identifications of AGN, but what process is actually producing the radio emission related to the AGN could be on pc to kpc scales. Although we explore this in the next sub-section, we will likely require ancillary information to help determine what the source of radio emission is in these $T_b$-identified AGN to make further progress.

\subsection{What is driving the radio emission?}

For HERGs and LERGs, the dominant radio emission mechanism is jets. But in radio-quiet AGN, we see that only 32 percent of the population are identified via their $T_b$. In these sources, we can separate the AGN and star formation luminosities (see Section~\ref{sec:sfagn}) and try to investigate the origin of the radio luminosity due to AGN. We consider three possibilities: star formation, winds, and jets. Two of these possibilities attribute the radio emission to AGN, and we include star formation as a comparison. The caveats in the previous section mean that our separation of radio emission from SF and AGN activity is not yet robust, and it is also possible that there is enhanced, compact star formation due to the AGN. For each of these three possibilities, we construct a distribution of radio luminosities and compare this with the distribution of radio AGN luminosities in the $T_b$-identified samples of unclassified, SFG, and RQAGN sources. 

For the star-formation distribution, we select all sources in the SFG class from Best et al., in prep. and remove the $T_b$-identified AGN. The radio luminosity in these sources is therefore dominated by star formation, and we simply use the integrated radio luminosity calculated from the standard-resolution image. Similarly for jets, we select all HERGs and LERGs and use their integrated radio luminosity. For winds, we rely on the models in \cite{nims_observational_2015}, which provide a prescription for estimating the synchrotron spectrum from the shocked ambient medium of an AGN wind. We use their Eqn. 32 to estimate the radio luminosity at 144 MHz from this shocked ambient medium. We assume the same fiducial model as \cite{nims_observational_2015}, but we require a realistic distribution of bolometric luminosities. These we take from \cite{woo_active_2002}, who compiled values from the literature and estimated values for another $\sim$200 galaxies. From this sample we select only the lower-luminosity Seyfert galaxies and radio-quiet AGN and use this distribution of bolometric luminosities to estimate the wind contribution to the radio luminosity.

\begin{figure}
	\centering
	\includegraphics[width=0.5\textwidth]{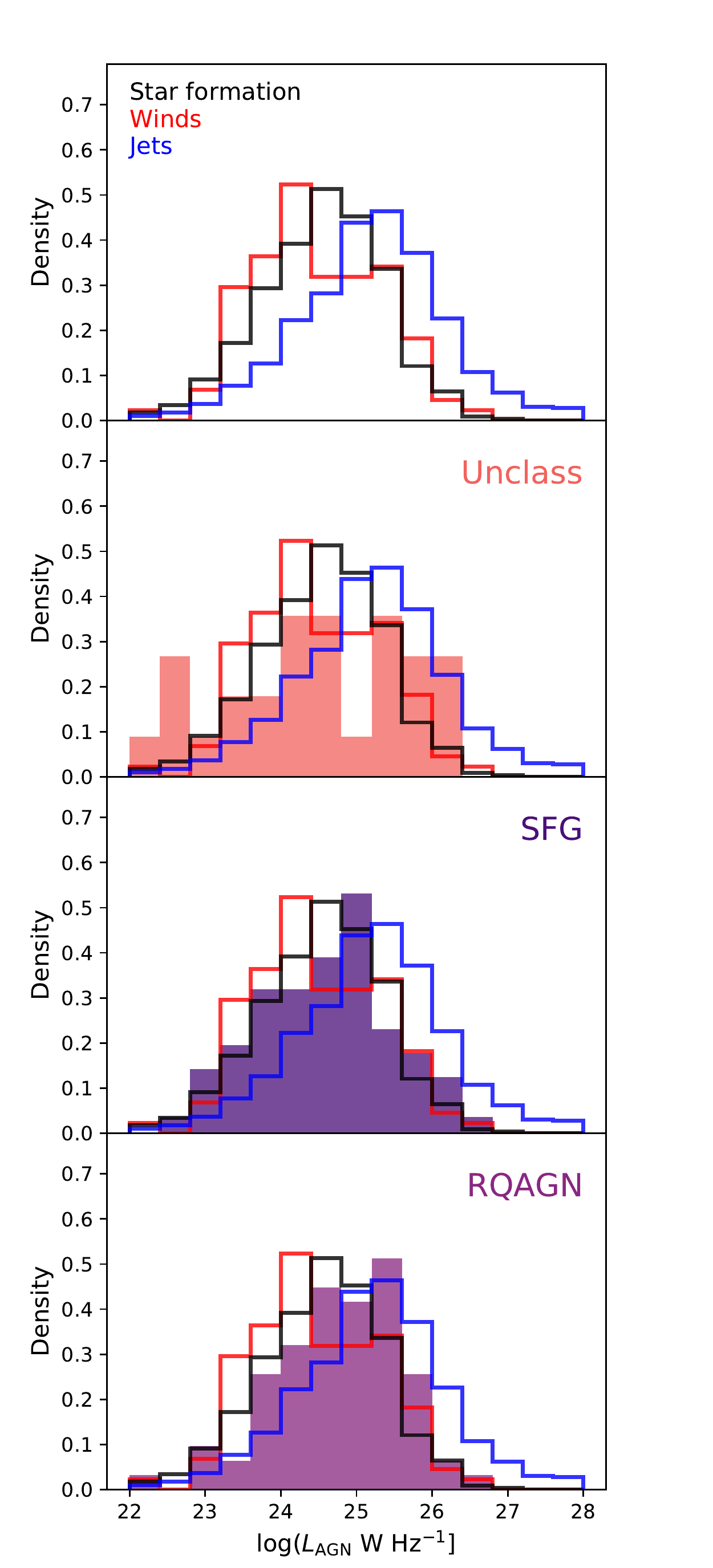}
	\caption{The top panel shows distributions for three possible contributions to AGN radio luminosity in radio-quiet AGN: star formation, winds, and jets. The following three panels show these distributions overlaid on the distributions of AGN luminosity for the $T_b$-identified AGN from the unclassified, SFG, and RQAGN samples.}
	\label{fig:sources}
\end{figure}

Figure~\ref{fig:sources} shows these three representative distributions in the top panel. The distributions of winds and star formation cover similar ranges, with only slight differences in the peaks of their distribution. The jet distribution is shifted towards higher radio luminosities, as expected. All three distributions are then overlaid on the distributions of AGN luminosity for the unclassified, SFG, and RQAGN samples, for a qualitative comparison. 

Both the SFG and RQAGN samples have distributions of $L_{\mathrm{AGN}}$ (calculated from the peak intensity in the high-resolution image) which fall within the distributions of all three potential sources of radio emission, while the unclassified sample has a slight over-abundance at the very low $L_{\mathrm{AGN}}$ end, also has peaks which seem to be consistent with the distributions of $L_{\mathrm{AGN}}$ from star-formation and jets, confirming their composite nature. It is possible that the unclassified sources have photometric redshifts very different from the actual redshifts, resulting in the inability of the SED fitting to properly classify them and yielding incorrect estimates of the AGN luminosity. Follow-up spectroscopic observations to secure redshifts for these sources will be critical for helping understand their nature. 

The SFG sample appears to have a single peak in its distribution at $L_{\mathrm{AGN}}\,\sim 10^{25}$ W$\,$Hz$^{-1}$.  We have seen that the SFG population has infrared colours on the borderline between AGN and SFGs, and the population has median $L_{\mathrm{AGN}} / L_{\mathrm{total}} = 0.44\pm 0.14$. The high-$T_b$ cores could be due truly to AGN activity, or intense starbursts in the nuclear region. However, Sweijen et al. (in preparation) examines a sample of hyper-luminous infrared galaxies (HyLIRGs) and finds that their brightness temperatures imply that 98 percent of the sample detected in the high-resolution image are likely to host radio AGN. The fact that the SFG population appears to have a single Gaussian distribution of $L_{\mathrm{AGN}}$ implies that the radio emission mechanism is similar across the population, and the fact that the distribution is shifted slightly more towards the star formation and winds distributions could mean that it is less likely these galaxies host jets. Spatially resolved IFU data may be needed to confirm the presence of AGN. 

The RQAGN sample looks like it could be two overlapping distributions, one peaking around $L_{\mathrm{AGN}}\,\sim 10^{24.5}$ W$\,$Hz$^{-1}$, and one around $\sim 10^{25.5}$ W$\,$Hz$^{-1}$. It is likely the population hosts a mix of different radio emission mechanisms, which is also supported by the fact that only 32 percent of the overall RQAGN population are identified as AGN via their $T_b$ values. There seems to be a significant fraction of the sample which is shifted towards higher $L_{\mathrm{AGN}}$, so it is likely at least some of these host radio jets. However, the lower peak could be due to either winds or extremely compact starbursts. \cite{nims_observational_2015} make predictions for the spectral index expected for radio emission from the shocked ambient medium due to winds, and future observations providing spectral information on the same spatial scales, e.g. with e-MERLIN, will be interesting to test this.

\subsection{Coevolution}

\begin{figure}
\centering
\includegraphics[width=0.5\textwidth]{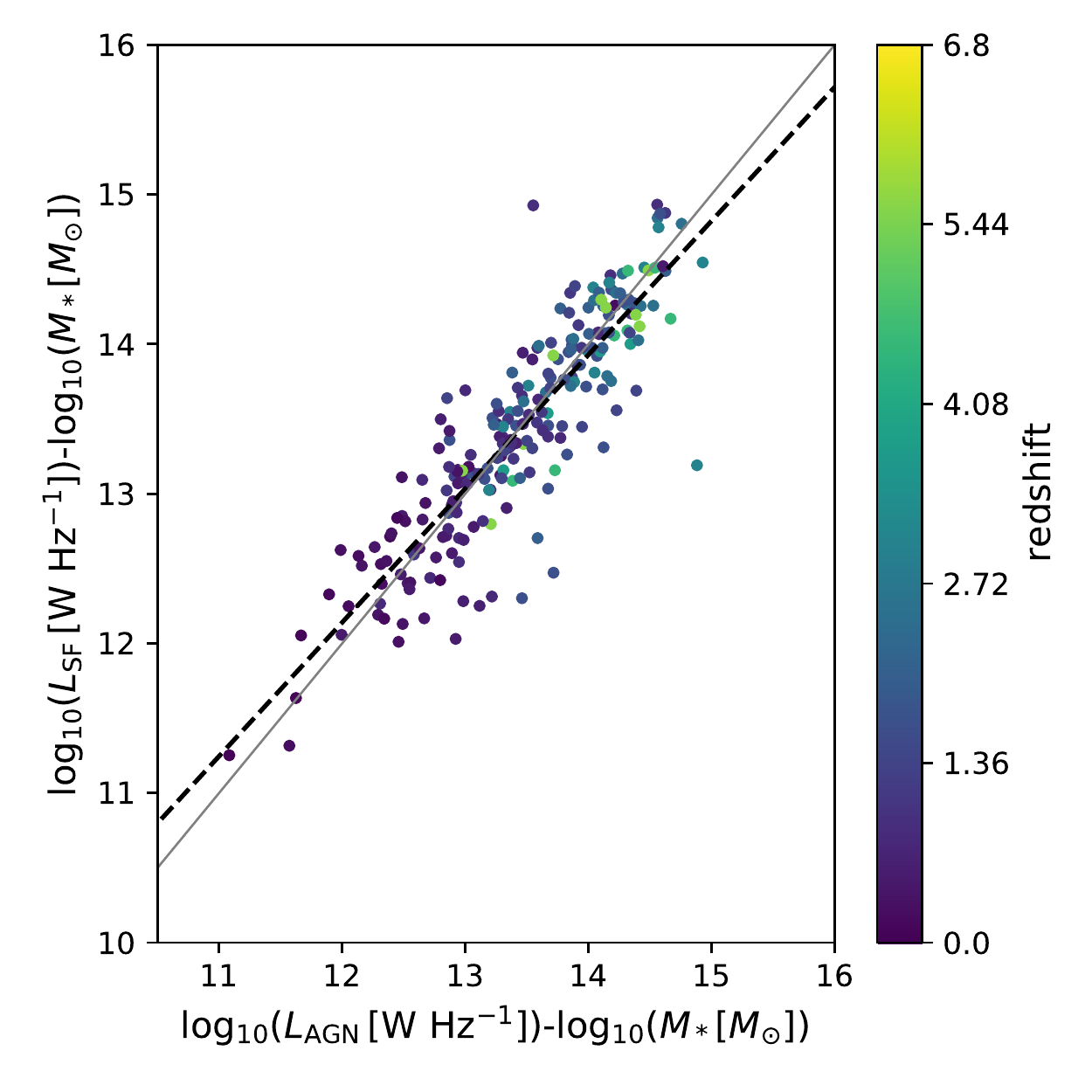}
\caption{The stellar-mass normalised $L_{\mathrm{SF}}$ versus the stellar-mass normalised $L_{\mathrm{AGN}}$ for unresolved, non-radio excess, $T_b$-identified AGN. The points are coloured by their redshift. The gray dashed line is a fit to the data, while the thin gray line shows the one-to-one line.}
\label{fig:coeval}
\end{figure}

Understanding the co-evolution of super-massive black holes and their host galaxies is linked to understanding the relative contributions of AGN activity and star formation to a galaxy's overall energy budget. Separating the radio emission from AGN and star formation as we did in Section~\ref{sec:sfagn} affords us the opportunity to see how they are linked, although correlation does not mean causation. Figure~\ref{fig:coeval} shows the luminosity from star formation vs. the luminosity from AGN for unresolved, non-radio excess, $T_b$-identified AGN. We normalise by the stellar mass to remove any redshift bias. A relation is clear, with a linear fit yielding: 
log$_{10}(L_{\textrm{SFR}}/M_*)= (0.89\pm0.03)\times$log$_{10}(L_{\textrm{AGN}}/M_*) + (1.41\pm0.43)$. The slope of the linear fit is slightly offset from unity. 

Colouring the points by their redshift reveals that this relationship may still have some dependence on redshift. When restricting the linear fit to include only the 29 sources in the sample which have spectroscopic redshifts, the fit changes to: log$_{10}(L_{\textrm{SFR}}/M_*)= (0.92\pm0.06)\times$log$_{10}(L_{\textrm{AGN}}/M_*) + (1.19\pm0.83)$. While the slope becomes slightly steeper, it is still consistent with the overall fit within the uncertainty on the parameter. 

Overall the tight correlation between the stellar-mass normalised radio luminosities due to star formation and AGN supports the idea that super-massive black holes co-evolve with their galaxies. The caveats outlined previously in this section mean the exact form of this relationship should be approached with caution, and this issue will be explored further in future studies.

\section{Conclusions and future work}
\label{sec:con}

In this paper we have demonstrated, for the first time, the use of brightness temperature measurements at low (144 MHz) radio frequencies to identify 940 AGN. This is only possible thanks to the sub-arcsecond resolution of the ILT, which is critical for reaching the limiting values of brightness temperature generated by a star-forming galaxy. We find that 89 percent of \tbpeak\ and 83 percent of \tbtotal\ selected sources are also classified as AGN via photometric or SED classification methods. This demonstrates that in the absence of detailed SED fitting (which provides the bulk of the AGN classifications, compared to photometric methods), we can reliably select AGN using $T_b$ at low frequencies. This will be crucial as we post-process the LOFAR Two-metre Sky Survey \citep[LoTSS;][]{shimwell_lofar_2022,shimwell_lofar_2019} at high resolution, as we will not have the exquisite ancillary data available in, e.g., the Lockman Hole region, for the entire Northern Sky. 

Investigating the ancillary data, we find that infrared colour-colour plots show SFGs with a $T_b$ AGN identification are either in or around the edges of the selection wedges, consistent with them being composite AGN/SFG sources. The majority of RQAGN lie within the selection wedges, while the unclassified sources split into two groups: one firmly within the AGN selection criteria, and one outside of it. 

Using both the rms map from the high-resolution image and the peak brightness from the standard-resolution image, there are 13,439 potentially detectable sources of which only 2,214 sources are detected\footnote{Specifically, there are 2,214 5$\sigma$ sources after removing duplicates and those with no redshift information.}. The fraction of detected sources increases up to $\sim$5 mJy, then is fairly constant at higher flux densities, while below this the high-$T_b$ population splits from the overall population, indicating that we are detecting a mixture of SFGs and AGN in the high-resolution image. 

Dividing the $T_b$-identified AGN sample into sub-populations, we find qualitatively similar behaviour as in Best et al., in prep. for the LERG and HERG populations. The high-$T_b$ sources comprise 57 and 68 percent of these populations, respectively. Over half (61 percent) of the unclassified sources are high-$T_b$ sources. The high-$T_b$ RQAGN only make up 32 percent of the overall population, indicating that there is likely a mixture of radio emission mechanisms in the overall RQAGN population. Finally, only 20 percent of SFGs are identified as AGN via their high-$T_b$; when considered in context of their placement on the infrared colour-colour plots these are highly likely to be composite AGN/SFG sources, consistent with the results of \cite{strazzullo_deep_2010} for the Deep Swire Field, or LERGs which fall short of the radio excess cut-off. This is in contrast to a sample of HyLIRGs, in which Sweijen et al. (in prep) find 98 percent of the sample to be $T_b$-identified AGN. 

With the simple assumption that the peak brightness of unresolved sources in the high-resolution map represents the AGN luminosity, we place the unclassified sources on the $L_R\,\sim\,$SFR relation and find that none have a radio excess. The ratio of AGN to total radio luminosity in all non-radio excess, unresolved sources ranges from less than 0.1 to almost unity, with a median value of 0.49$\pm$0.16, which again suggests composite systems. For RQAGN, this fraction is the closest to being correlated with $f_{\mathrm{AGN}}$ from the SED fitting, but the other populations show weaker correlations. We find only a very weak dependence on $L_R$, with large uncertainties. 

This first step in using the unique observations from \cite{sweijen_deep_2022} shows the potential for using the ILT as a tool for identifying AGN based on their radio data alone, although there is more work to be done to understand these samples of AGN. The new WEAVE multi-object survey spectrograph on the William Herschel Telescope will collect spectroscopic redshifts for the sample outlined here as part of the WEAVE-LOFAR survey \citep{smith_weave-lofar_2016}, allowing us to refine the brightness temperature measurements. Complementary radio data on the same spatial scales but at GHz frequencies from an EVN/e-MERLIN project will help us understand their spectral properties (PI: McKean). In a future paper, we will use a forward-modelling approach to better constrain the separation of star-formation and AGN activity. This will allow us to fully exploit these new sub-arcsecond, low-frequency datasets to understand how AGN co-evolve with their galaxies.

\section*{Acknowledgements}
The authors acknowledge the contributions of S.R. Ward, who worked on a related project that helped kickstart this one, and J.C.S. Pierce, who helped with early visual identifications of sources. 
This work was supported by the Medical Research Council [MR/T042842/1]. 
PNB is grateful for support from the UK STFC via grant ST/V000594/1. 
KJD acknowledges funding from the European Union’s Horizon 2020 research and innovation programme under the Marie Sk\l{}odowska-Curie grant agreement No. 892117 (HIZRAD). 
IP acknowledges support from INAF under the SKA/CTA PRIN “FORECaST” and the PRIN MAIN STREAM “SAuROS” projects. 
WLW  acknowledges support from the CAS-NWO programme for radio astronomy with project number 629.001.024, which is financed by the Netherlands Organisation for Scientific Research (NWO). 
RJvW acknowledges support from the ERC Starting Grant ClusterWeb 804208. 
This paper is based (in part) on data obtained with the International LOFAR Telescope (ILT) under project code LT10\_012. LOFAR \citep{van_haarlem_lofar:_2013} is the Low Frequency Array designed and constructed by ASTRON. It has observing, data processing, and data storage facilities in several countries, that are owned by various parties (each with their own funding sources), and that are collectively operated by the ILT foundation under a joint scientific policy. The ILT resources have benefited from the following recent major funding sources: CNRS-INSU, Observatoire de Paris and Université d’Orléans, France; BMBF, MIWF-NRW, MPG, Germany; Science Foundation Ireland (SFI), Department of Business, Enterprise and Innovation (DBEI), Ireland; NWO, The Netherlands; The Science and Technology Facilities Council, UK; Ministry of Science and Higher Education, Poland. This work has made use of the Dutch national e-infrastructure with the support of SURF Cooperative through grant e-infra 180169.

\section*{Data Availability}
The data used in this paper comes from \cite{sweijen_deep_2022} and the LoTSS Deep Fields Data Release 1, for which the radio data are presented in \cite{tasse_lofar_2021} and made publicly available through both the Centre de Données astronomiques de Strasbourg (CDS) and through the LOFAR Surveys website (\url{https://lofar-surveys.org/deepfields.html}). Multi-wavelength photometric catalogues and photometric redshifts come from \cite{kondapally_lofar_2021} and \cite{duncan_lofar_2021} respectively, both of which are also available through CDS and the LOFAR Surveys website. The radio data from \cite{sweijen_deep_2022} is also available on the LOFAR Surveys website (\url{https://lofar-surveys.org/hdfields.html}) and we make additional use of the deconvolved sizes which are available upon request. This data was supplemented with SED fitting results presented in Best et al., in prep. which are available on request and will be published with the Best et al paper, through \url{https://lofar-surveys.org/deepfields.html}. Supporting information was taken from tables published in \cite{middelberg_mosaiced_2013} and \cite{woo_active_2002}, and are available in csv format at \url{https://github.com/lmorabit/AGNdetect}, where the publicly available code for the analysis presented here can also be found.

\bibliographystyle{mnras}
\bibliography{references.bib}

\appendix
\section{Comparison with VLBA observations}
\label{app:vlba}

We can compare our results with a mosaicked wide-field VLBI study of a portion of the Lockman Hole field \citep[][project code BM332]{middelberg_mosaiced_2013}. Although this area only covers 0.59$\,$deg$^2$ with three overlapping pointing centres\footnote{from A. Deller, private communication: \\ Epochs A/D: 10h52m56.0s +57d29'06.0" \\ Epoch B: 10h52m08.8s 57d21'33.8" \\ Epoch C: 10h51m02.0s +57d13'50.4"} rather than the 6.6$\,$deg$^2$ from \cite{sweijen_deep_2022}, they do overlap which allows for a direct comparison. The \cite{middelberg_mosaiced_2013} study used the Very Long Baseline Array (VLBA) at 1.3 GHz, targetting 496 sources with integrated flux densities $>100\,\mu$Jy, as measured with the Very Large Array (VLA). After calibration, the VLA peak brightness was compared to the local rms in the VLBI maps, and only 217 sources were deemed to be potentially detectable. Of these, there were 65 detections with 6$\sigma$ significance. Note that at this frequency, any detection with the VLBA is considered to satisfy the $T_b$ criteria for AGN identification. The $T_b$-identified AGN sample therefore implies a sky density of 109.96 sources per deg$^2$. Within the same region of sky, there are 92 $T_b$-identified AGN from the high-resolution image, yielding a sky density of 155.64 sources per deg$^2$. These numbers are approximately similar, although a direct comparison is difficult because the two studies are conducted at different frequencies. The spectral indices of the VLBA sample, calculated using VLA and GMRT observations,  are diverse, which means scaling the sensitivity from 1.3 GHz to 144 MHz using a single spectral index is not reliable for predicting the outcome. 

\begin{figure}
	\centering
	\includegraphics[width=0.45\textwidth]{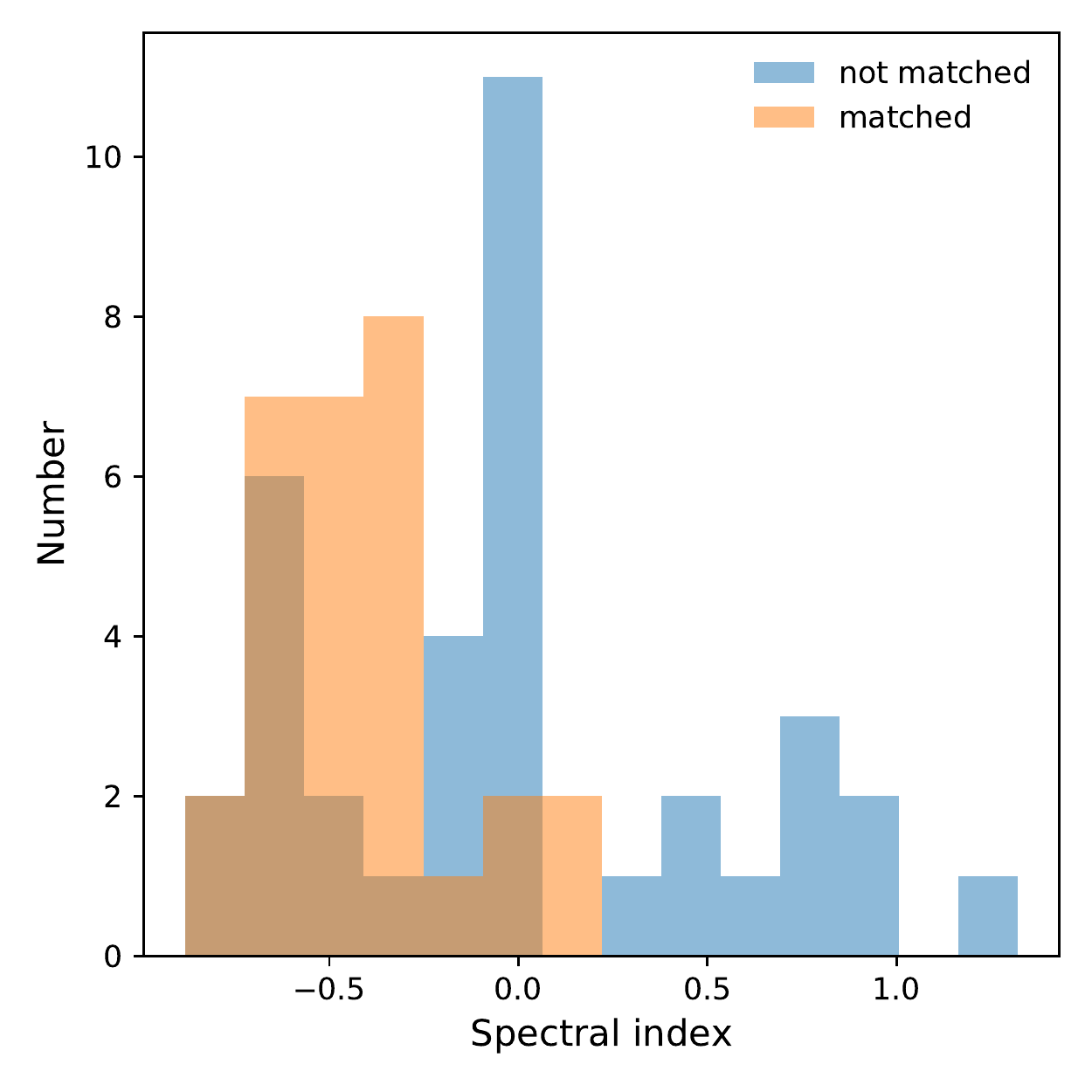}
	\includegraphics[width=0.45\textwidth]{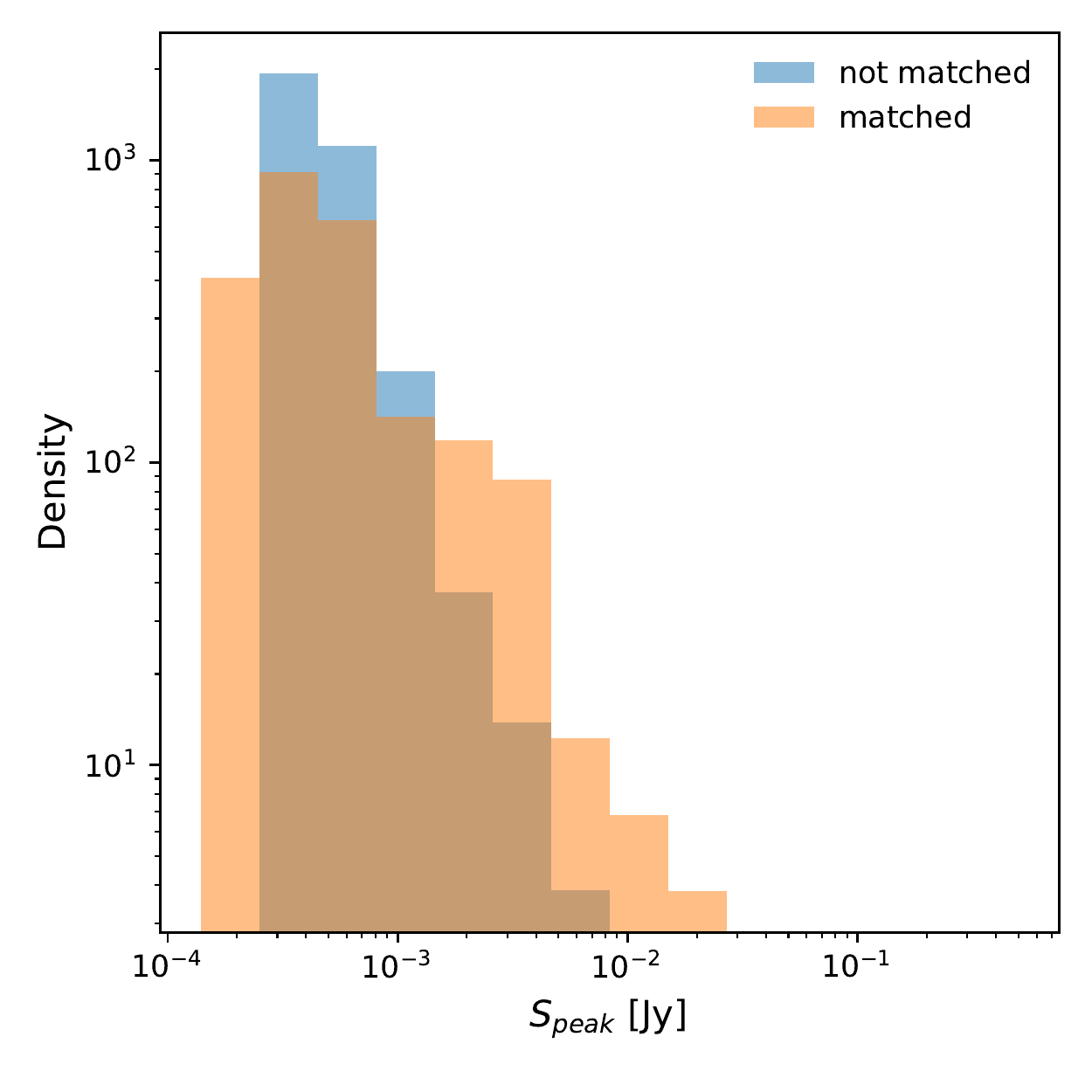}
	\caption{\textit{Top:} Spectral index distributions of the VLBA sources which are matched (orange) and not matched (blue) to the $T_b$-identified sources in this study. \textit{Bottom:} Peak brightness from the high-resolution image for sources which are matched (orange) and not matched (blue) to the VLBA sources.}
	\label{fig:valpha}
\end{figure}

Cross-matching the $T_b$-identified samples at MHz and GHz frequencies, we find only 22 matches (15 \tbpeak\ sources, and 7 \tbtotal\ sources). That means there are 43 unique VLBA detections, and 70 unique $T_b$-identified AGN at 144 MHz. Using the VLA to GMRT spectral indices for the VLBA sources, the distribution of which is shown in Fig.~\ref{fig:valpha}, we can see that those which are not matched are predominantly flat-spectrum sources. However, it is worth keeping in mind that the spectral indices were calculated from lower-resolution data, and may not reflect the spectral index of the compact AGN component. Looking at the peak brightness at 144 MHz we can see (Fig.~\ref{fig:valpha}) that the VLBA-matched sources are, on average, brighter. 

One must consider that although both studies use $T_b$ as an AGN identification tool, the beam solid angles at these different frequencies are an order of magnitude different, and therefore will be probing different physical scales. Future work exploiting the EVN and e-MERLIN will provide matched resolution to the high-resolution image studied here, and will be more appropriate for drawing meaningful conclusions.

\label{lastpage}

\end{document}